\documentclass[aps,pra,epsf,superscriptaddress,amsmath,amssymb,amsfonts,twocolumn,showpacs,floatfix]{revtex4-1}

\usepackage[utf8]{inputenc}

\usepackage{amsmath}
\usepackage{amssymb}
\usepackage{amsfonts}
\usepackage{amstext}
\usepackage{amsthm}
\usepackage{mathtools}
\usepackage{physics}

\usepackage{bm} 

\usepackage{graphicx}
\graphicspath{{figures/}}

\usepackage{epsfig}
\usepackage{dcolumn}
\usepackage{bm}
\usepackage{braket}
\usepackage{color}
\usepackage[colorlinks=true,citecolor=blue]{hyperref}

\usepackage[english]{babel}

\usepackage{multirow}

\usepackage[dvipsnames]{xcolor}

\definecolor{deeppurple}{rgb}{0.7, 0, 0.8}

\allowdisplaybreaks
\advance\parskip 0.4pt

\usepackage{hyperref}
\hypersetup{      
	urlcolor=blue
}

\usepackage{natbib}

\begin{document}

\title{Theoretical and numerical evidence for the potential realization of the Peregrine soliton  in repulsive two-component Bose-Einstein condensates}

\author{A. Romero-Ros}
\affiliation{Center for Optical Quantum Technologies, Department of Physics,
University of Hamburg, Luruper Chaussee 149, 22761 Hamburg,
Germany}

\author{G. C. Katsimiga}
\affiliation{Center for Optical Quantum Technologies, Department of Physics,
University of Hamburg, Luruper Chaussee 149, 22761 Hamburg,
Germany}
\affiliation{The Hamburg Centre for Ultrafast Imaging,
University of Hamburg, Luruper Chaussee 149, 22761 Hamburg,
Germany}

\author{S. I. Mistakidis}
\affiliation{ITAMP, Center for Astrophysics $|$ Harvard $\&$ Smithsonian, Cambridge, MA 02138 USA}
\affiliation{Department of Physics, Harvard University, Cambridge, Massachusetts 02138, USA}

\author{B.~Prinari}
\affiliation{Department of Mathematics, State University
	of New York, Buffalo, New York 14260, USA}

\author{G. Biondini}
\affiliation{Department of Mathematics, State University of New York, Buffalo, New York 14260, USA}
\affiliation{Department of Physics, State University of New York, Buffalo, New York 14260, USA}

\author{P. Schmelcher}
\affiliation{Center for Optical Quantum Technologies, Department of Physics,
University of Hamburg, Luruper Chaussee 149, 22761 Hamburg,
Germany}
\affiliation{The Hamburg Centre for Ultrafast Imaging,
University of Hamburg, Luruper Chaussee 149, 22761 Hamburg,
Germany}

\author{P. G. Kevrekidis}
\affiliation{Department of Mathematics and Statistics, University
of Massachusetts Amherst, Amherst, MA 01003-4515, USA}
\date{\today}

\begin{abstract}
The present work is motivated by the recent experimental realization of the Townes soliton in an effective two-component Bose-Einstein condensate by B.~Bakkali-Hassan {\it et al.} \href{https://journals.aps.org/prl/abstract/10.1103/PhysRevLett.127.023603}{[Phys. Rev. Lett. {\bf127}, 023603 (2021)]}.
Here, we use a similar multicomponent platform to exemplify theoretically
and numerically, within the mean-field Gross-Pitaevskii framework, the potential toward the experimental realization of a different fundamental wave structure, namely the Peregrine soliton.
Leveraging the effective attractive interaction produced within the mixture's minority species in the immiscible regime, we illustrate how initialization of the condensate with a suitable power-law decaying spatial density pattern yields the robust emergence of the Peregrine wave in the absence and in the presence of a parabolic trap. 
We then showcase the spontaneous emergence of the Peregrine soliton via a suitably crafted wide Gaussian initialization, again both in the homogeneous case and in the trap scenario.
It is also found that narrower wave packets may result in periodic revivals of the Peregrine soliton, while broader ones give rise to a cascade of Peregrine solitons arranged in a so-called Christmas-tree structure. 
Strikingly, the persistence of these rogue-wave structures is demonstrated in certain temperature regimes as well as in the presence of transversal excitations through three-dimensional computations in a quasi-one-dimensional regime.
This proof-of-principle illustration is expected to represent a practically feasible way to generate and observe this rogue wave in realistic current ultracold atom experimental settings.
\end{abstract}

\maketitle

\section{Introduction} \label{sec:introduction}

In 1966, Draper {\it et al}.~\cite{Draper1966} reported the detection of an oceanic wave event featuring a \textit{freak} wave, namely a wave several times bigger than the average {sea state}.
Nowadays these freak waves are referred to as rogue waves~\cite{Kharif2009}.
Rogue waves are extreme wave events that emerge out of nowhere and disappear without a trace~\cite{Akhmediev2009,Ablowitz2021}.
Under appropriate approximations, they can be mathematically described by {solutions} of the nonlinear Schr\"odinger equation (NLS)~\cite{Kuznetsov1977,Ma1979,Dysthe1999,Akhmediev1986,Karjanto2020}.
This mathematical description of rogue waves allowed extrapolation of these phenomena to a large variety of nonlinear physical systems, other than oceanic waves, ranging from nonlinear optics~\cite{Kibler2010,Kibler2012,DeVore2013,Dudley2014,Frisquet2016} to plasmas~\cite{Sabry2012,Bains2014,Tolba2015}, and from liquid helium~\cite{Ganshin2008} to Bose-Einstein condensates (BECs)~\cite{Charalampidis2018,Bakkali-Hassani2021} (see also the reviews of Refs.~\cite{Yan2012a,Onorato2013,Mihalache2017}).

Among the different members of the rogue-wave family, arguably, the most celebrated one is the rational solution known as the Peregrine soliton~\cite{Peregrine1983}.
Contrary to the Kuznetsov-Ma soliton~\cite{Kuznetsov1977,Ma1979}, which is periodic in time, or the Akhmediev breather~\cite{Akhmediev1986}, which is periodic in space, the Peregrine soliton is a wave localized both in time and in space.
In recent years, Peregrine solitons have been successfully realized in water tank experiments~\cite{Chabchoub2011,Chabchoub2012,Chabchoub2014,Tikan2021},
plasmas~\cite{Bailung2011} and optical fibers~\cite{Kibler2010,Kibler2012,Tikan2017,Marcucci2019}, 
demonstrating the active interest of distinct communities in these, as well as similar wave events of higher order~\cite{Chowdury2021}.
All of the above-mentioned physical settings, however, involve self-focusing media.
Here, we report on the theoretical formulation and numerical implementation of the spontaneous nucleation of the Peregrine soliton in self-defocusing media within the mean-field framework.

Within the ultracold superfluid realm, scalar and multicomponent BECs  in the mean-field framework are accurately described by a variant of the NLS equation, the well-known Gross-Pitaevskii equation (GPE)~\cite{Pitaevskii2003}. 
In that light, it is natural to expect that rogue waves can exist in BEC systems~\cite{Zhao2013a,Li2018,Charalampidis2018,ChaachouaSameut2020,Bakkali-Hassani2021}. 
Importantly, the high degree of controllability of such settings, e.g., in terms of tunable interatomic interactions through the aid of Feshbach~\cite{Kohler2006,Chin2010} or confinement-induced~\cite{Olshanii1998,Kim2006} resonances, as well as the flexibility to realize almost arbitrary potential landscapes~\cite{Henderson2009,Grimm2000},
renders these platforms ideal test beds for the study of rogue-wave formation.

It is also in this BEC context that the recent experimental realization of the so-called Townes soliton~\cite{Chiao1964SelfTrapping} came to fruition~\cite{Bakkali-Hassani2021}. The Townes
soliton is a planar, real, nodeless and radially symmetric stationary solution of the two-dimensional one-component {\it focusing} (i.e., attractive interaction) GPE. 
Interestingly, the Townes soliton was conceived theoretically~\cite{Chiao1964SelfTrapping}, and realized experimentally~\cite{Bakkali-Hassani2021} by reducing a two-component {\it defocusing} (i.e., repulsive interaction) setting to an effective \textit{focusing} single-component one for a minority component~\cite{Dutton2005} (in the presence of a dominant majority component~\footnote{It is relevant to note as an aside here that the Townes soliton has also been created via a different method very recently in~\cite{Chen2021}; the latter method involved the modulational instability of a bright solitonic stripe in two-dimensional space.}).
Motivated by this recent realization of an effectively attractive dynamics, manifested
in the broadly experimentally accessible two-component repulsive BEC setting~\cite{Pitaevskii2003,Kevrekidis2015}, and the earlier work of Ref.~\cite{Dutton2005}, 
here, we utilize this multicomponent platform to study the formation of the Peregrine soliton by solving the underlying GPEs. 
Recall that the realization of this rogue wave, which is an exact solution of the focusing NLS equation~\cite{Peregrine1983}, in the BEC context, remains until now an experimental challenge, at least in part due to the modulational instability which the background of this wave suffers~\cite{Ablowitz2021}. 
Here, we propose an alternative route for achieving the nucleation of this rational solution, paving the way for its controllable experimental observation in a repulsive BEC environment which can be routinely and stably produced 
in the laboratory. 

More concretely, we initially exemplify how, under the appropriate choice of the inter- and intra-component interactions, a two-component repulsive BEC can be effectively reduced to an attractive single-component one~\cite{Dutton2005,Bakkali-Hassani2021}, thus allowing for the spontaneous emergence of the Peregrine soliton. 
We investigate such a rogue-wave generation both in the absence and in the presence of an external harmonic trapping potential but also within the two- and the effective single-component model.
Here, very good agreement between the two models is demonstrated, verifying the existence of the Peregrine wave for the homogeneous setting while unveiling its recurrence in the confined setup.

We then study the nucleation of the rogue-wave pattern in the so-called semiclassical limit. 
The latter is addressed by initializing an experimentally accessible sufficiently wide Gaussian wavepacket.
It is found that the width of the Gaussian directly impacts the resulting dynamics, from the periodic revival of the Peregrine soliton toward the so-called umbilical gradient catastrophe~\cite{Bertola2013}. 
The latter leads, in turn, to the formation of a cascade of Peregrine waves, also referred to as a Christmas-tree pattern.
In our numerical computations, a Christmas-tree configuration is found to decay in the confined geometry, emitting dark-bright soliton-type structures which oscillate inside the parabolic trap featuring  unexpected trajectories. 
For a recent experiment on the controllable generation and current state-of-the-art on multicomponent dark-bright solitons, see, e.g., Ref.~\cite{Lannig2020Collisions}.
Furthermore, in order to expose the robust features of the Peregrine wave, we also explore a number of variations of the two-component setup. 
Namely, by considering mass-imbalanced mixtures we unveil that Peregrine formation, in general, can take place only in mixtures where the minority component is the heaviest one.
Specifically, in this mass-imbalanced situation, the dark-bright patterns experience a breathing motion, on top of their in-trap oscillation, whose frequency depends on the size of the solitons formed. 

Additionally, the robustness of the Peregrine in certain temperature regimes is unveiled through the dissipative Gross-Pitaevskii framework~\cite{Proukakis2008,Katsimiga2021Phase}. 
Last, but definitely not least, the spontaneous nucleation of the Peregrine and Christmas tree configurations is showcased in genuinely three-dimensional (3D) computations featuring a quasi-one-dimensional (1D) geometry. We believe that the numerical persistence of the configuration in this setting is strongly suggestive of the feasibility of our proposed experimental realization.

The flow of our presentation is as follows.
In Sec.~\ref{sec:model} we provide a description of the repulsive two-component setup along with its reduction to an effective single-component attractive model.
In Sec.~\ref{sec:results}, we elaborate on the dynamical generation and features of the Peregrine soliton.
First, we show the emergence of the Peregrine soliton in the absence and in the presence of an external trapping potential. 
Then, we extend our considerations to the semiclassical setting
by using as an initial condition a broad Gaussian wavepacket,
again without and with a trap.
In this latter scenario, we also consider the impact of mass-imbalance.
The effect of temperature on the nucleation of the Peregrine soliton is subsequently discussed in Sec.~\ref{imperfections_dyn}. 
Importantly, Sec.~\ref{quasi1D_dyn} elaborates on the generation of the Peregrine in 3D geometries yet in the realm of the quasi-1D regime.
Finally, Sec.~\ref{sec:conclusions} provides our conclusions and future perspectives.
Appendix~\ref{app:wedge-structure} presents a modulational instability structure that arises at long time dynamics, recently described and found in Refs.~\cite{Biondini2016b,Kraych2019} but, contrary to our setup, for focusing media.

\section{The Peregrine wave in a two-component setting} \label{sec:model}

\subsection{Mean-field description}

To emulate spontaneous Peregrine soliton 
generation, our starting point and the primary focus of
our considerations will be the zero temperature limit. 
The impact of dissipation is appreciated later on in 
Sec.~\ref{imperfections_dyn}.
In the aforementioned limit, and also in 1D, the wave 
functions obey the following dimensionless system of 
coupled GPEs~\cite{Pethick2008,Pitaevskii2003,Kevrekidis2015}:
\begin{subequations}
\label{eq:CGPE}
    \begin{align}
    i u_t = -\frac{1}{2\tilde{m}_u} u_{xx} + V_u(x) u + (g_{11} |u|^2 + g_{12} |v|^2) u \,,
    \label{eq:CGPE_u}
    \\
    i v_t = -\frac{1}{2\tilde{m}_v} v_{xx} + V_v(x) v + (g_{21} |u|^2 + g_{22} |v|^2) v \,,
    \label{eq:CGPE_v}
    \end{align}
\end{subequations}
The subscripts $t$ and $x$ indicate the time and spatial derivatives, respectively.
The field $u$ ($v$) describes the wave function of the majority (minority) component.
Additionally, $\tilde{m}_u=m_u/M$, $\tilde{m}_v=m_v/M$ where $m_u$, $m_v$ denote the mass of the corresponding component and $M=m_u m_v/(m_u+m_v)$ is the reduced mass.
For our demonstration we will assume the scenario $m_u=m_v\equiv m=1$, unless stated otherwise.
Such a mass-balanced mixture could be realized, for example, by the two different hyperfine states of $^{87}$Rb~\cite{Egorov2013} (see also the discussion below).
The external potential of the system is given in the familiar parabolic form $V_j(x)=\frac{1}{2}\tilde{m}_j\Omega^2 x^2$.
Here $\Omega=\omega_x/\omega_{\perp}$, where $\omega_x$ and $\omega_{\perp}$ denote the longitudinal and transverse trapping frequencies of the system, respectively.
For a 1D cigar-shaped trap they should obey $\omega_x \ll \omega_{\perp}$. 
Also, $g_{jj} = \frac{1}{ \tilde{m}_j}\frac{a_{jj}}{a_\perp}$ and $g_{jk} = \frac{1}{2}\frac{a_{jk}}{a_\perp}$ are the effective intra- and inter-component 1D interaction strengths, respectively, with $a_{jk}$ denoting the $s$-wave scattering lengths accounting for collisions between atoms of the same ($j = k$) or different ($j\neq k$) species.

In the dimensionless units used here, the  densities $|u|^2$ and $|v|^2$, length, energy, and 
time are measured in units of $(2a_\perp)^{-1}$, $a_\perp=\sqrt{\hbar/(M \omega_\perp)}$, $\hbar \omega_\perp$, and $\omega_\perp^{-1}$, respectively. 
In this sense, typical evolution times of the order of $10^3$ when considering, for instance, a trap with $\omega_x\approx 2 \pi \times 1\,$Hz and $\omega_{\perp}= 2 \pi \times 400\,$Hz correspond to $\sim400\,$ms, see also Section~\ref{quasi1D_dyn}.

\subsection{Reduction to a single-component model} 

The key feature of our analysis lies in considering the limit where $|g_{12}-g_{11}| \ll g_{11}$, as well as $|g_{22}-g_{11}| \ll g_{11}$.
In this setup, an effective single-component description of the two-component system can be achieved assuming that one component is effectively immersed in a bath of atoms of the second one~\cite{Dutton2005,Bakkali-Hassani2021}.  
Notice that similar considerations~\cite{Johnson2012Breathing,Mistakidis2019a,Mistakidis2021a} are also utilized in rather distinct contexts such as the dressing of impurities by the excitations of a many-body medium, leading to the concept of polarons~\cite{Massignan2014Polarons}.
This scheme can be implemented experimentally via a two-photon Raman transition, where a transfer of a fraction of atoms (in wave functions of different types -- for details, see below) from the majority component will be made to the minority species.
Importantly, this allows us to keep the total density constant. 
In line with the recent experimental description of Ref.~\cite{Bakkali-Hassani2021}, our assumption will be that the two species add up to a Thomas-Fermi profile (since the chemical potential $\mu_u \gg \Omega$ will be used in the majority species) when $\Omega \neq 0$.
In the case of $\Omega=0$ (which will be employed first in order to showcase the ideas in a uniform setup), the total density of the two species is a constant background.

Then, following the above considerations, the dynamics of the minority species in the two-component system can be described by an effective single-component GPE given by~\cite{Dutton2005}:
\begin{eqnarray}
    i \partial_t v_\textrm{eff}=-\frac{1}{2} \partial_x^2 v_\textrm{eff} + V(x) v_\textrm{eff} + g |v_\textrm{eff}|^2 v_\textrm{eff}.
    \label{eq:eff_CGPE}
\end{eqnarray}
The key feature of this description is that the effective nonlinearity parameter $g$ here reads:
\begin{eqnarray}
    g=g_{22} - \frac{g_{12}^2}{g_{11}}.
\label{eq:eff_g}
\end{eqnarray}
As a result, if the condensates are on the (weakly) immiscible side, as is the case for $^{87}$Rb hyperfine states, we expect the effective nonlinearity to be attractive for our effective single-component species $v_\textrm{eff}$ that is approximately equal to $v$.
Here, motivated by relevant studies, such as those of Ref.~\cite{Mertes2007} and, more recently, Ref.~\cite{Egorov2013}, we will consider $g_{11}=1.004$ and $g_{22}=0.95$.
The value of $g_{12}$ is considered to be close to $0.98$; however, in order to enable the relevant (weak) immiscibility effect to be amplified and be visible at shorter timescales, here we will assume that $g_{12}=1.1$.
Effectively, it is well known that one of the scattering lengths can be tuned via techniques such as Feshbach resonance over wide parametric windows~\cite{Pollack2009}.
In that light, the relevant phenomenology should be observable, for example, in the hyperfine states $|1,-1\rangle$ and $|2,1\rangle$ of $^{87}$Rb for which the above parameters are given.
Besides, the tunability to different $g_{12}$ as used here is, in principle, accessible.
It is relevant to note also that it is not central to our considerations that $g_{12}$ is tuned.
Indeed, the results presented herein will be valid for the minority component more
generally within the immiscible regime, as the latter leads to the negativity of the expression of Eq.~(\ref{eq:eff_g}) for $g$ and, hence, the attractive nature of the effective
one-component description considered. 

\subsection{Peregrine ansatz and computational setup} \label{sec:setup}

Having set up this effectively attractive interaction, it is then relevant to discuss the coherent structure of interest, namely the Peregrine soliton, as a prototypical member of the family of rogue waves.
The relevant solution of Eq.~\eqref{eq:eff_CGPE} with $V(x)=0$ reads~\cite{Peregrine1983}:
\begin{eqnarray}
v_\textrm{eff}(x,t)=\sqrt{P_o}
\left[1 - \frac{4 \left(1 + 2 i \frac{t-t_o}{T_P}\right)}{1+4 (\frac{x-x_o}{L_P})^2 + 4
(\frac{t-t_o}{T_P})^2}\right] e^{i \frac{t-t_o}{T_P}}.
\label{eq:peregrine}
\end{eqnarray}
Here, $T_P=L_P^2=1/(g P_o)$ represents the characteristic scales of time and space of the density variation of the solution, respectively, while $P_o$ represents the background density of the minority component.
This implies that this is a monoparametric family of solutions, i.e., once $P_o$ is set, so are $T_P$ and $L_P$. 
Moreover, $t_o$ and $x_o$ denote, respectively, the time instant and location at which the Peregrine soliton emerges.
In what follows, we set $x_o=0$, unless stated otherwise.

Equation \eqref{eq:peregrine} is the solution that we will seek to effectively realize in our investigation in two distinct ways.
The first one, which involves the proof of principle, will consist of the initialization of the Peregrine wave form and the monitoring of its time-evolution.
It is assumed, therefore, that such a profile is transferred to the minority component at a substantially lower density $P_o$ than that of the $\mathcal{O}(1)$ 
majority component (i.e., $P_o \ll \mu_u$), while the majority component represents the remainder of the background toward a cumulative density of either constant value when $V(x)=0$, or a Thomas-Fermi cloud when $V(x) \neq 0$.
By Thomas-Fermi cloud here, we mean the ground-state profile in our setting of large chemical potential $\mu_u$.
As mentioned above, we perform two types of investigations, one without an external trapping potential and one in the presence of the parabolic trap.
Furthermore, in each of the examples, we perform two complementary explorations.
In the first one, we simulate the full two-component system of Eqs.~\eqref{eq:CGPE_u} and \eqref{eq:CGPE_v}, while in the second one we restrict our considerations to the one-component effective system of Eq.~\eqref{eq:eff_CGPE}. 
Our scope is to illustrate the relevance of the reduction of the former to the latter and to identify the case examples where this reduction may fail.

In addition to the proof-of-principle demonstration that an initialization of the Peregrine initial condition (well before its formation) will indeed lead to its emergence, we also want to {address} a more practical question.
In particular, it is straightforward to appreciate that the slowly {decaying} spatial wave form 
of Eq.~\eqref{eq:peregrine} (to a constant intensity background $P_o$, no less) may be more difficult to achieve in practice.
Hence, it is natural to seek a ``generic'' wave form that may lead to such an emergence, upon a straightforward initialization, e.g., with a Gaussian profile.
Here, we leverage the earlier findings of the rigorous work of Ref.~\cite{Bertola2013} in the integrable NLS setting, within the so-called semiclassical regime.
For our purposes, practically, this concerns wave functions with sufficiently large spatial width.
In this context, the authors of Ref.~\cite{Bertola2013} have identified a generic so-called umbilical gradient catastrophe which leads to the formation of a cascade of Peregrine waves, a structure that has been referred to as Christmas-tree in Ref.~\cite{Charalampidis2018}
and has recently been identified also in single and coupled phononic crystals~\cite{Charalampidis2018a,Miyazawa2021}. 
The relevant wave structures emerge at the poles of the so-called tritronqu{\'e}e solution of the Painlevé~I equation.
The principal result for our purposes is that the Peregrine and the Christmas-tree structures should emerge {\it spontaneously} from quite generic (wide, and thus effectively semiclassical) wave forms, such as a Gaussian, but also sech-shaped ones and others~\cite{Biondini2020Semiclassical}; indeed, the key feature
is the width of the localized wave form, rather than its 
concrete functional form.
Specifically, it has been demonstrated that the occurrence of the Christmas-tree structure is rather universal, in the sense that it appears as a result of strong modulational instability from different initial configurations~\cite{Biondini2020Semiclassical}.
This motivates the second set of our numerical experiments where, instead of initializing a precise Peregrine, we exploit a broad Gaussian of the form $v=A \exp(-x^2/w^2)$ and observe the resulting evolution. 
Typical values of the Gaussian amplitude $A=0.2$ and width $w=50, 150$ are used in the results below~\footnote{For a transverse confinement with $\omega_{\perp}=2\pi \times150\,$Hz these widths refer to a range from $45\,\mu$m to $135\,\mu$m.}. 
For the numerical investigations that follow, we use a fourth order Runge-Kutta integrator with spatial and temporal discretization steps 
$dx=0.1$ and $dt=0.001$, respectively. 
Additionally, the system size for the homogeneous settings is $[-3000, 3000]$, in the dimensionless units adopted herein, while for the trapped studies it is $\big[-\frac{3}{2}r_\textrm{TF}, \frac{3}{2}r_\textrm{TF}\big]$. 
Here, $r_\textrm{TF}=\sqrt{2\mu_u}/\Omega$ denotes the Thomas-Fermi radius of the majority $u$ component.

\begin{figure}[t]
    \centering
    \includegraphics[width=\linewidth]{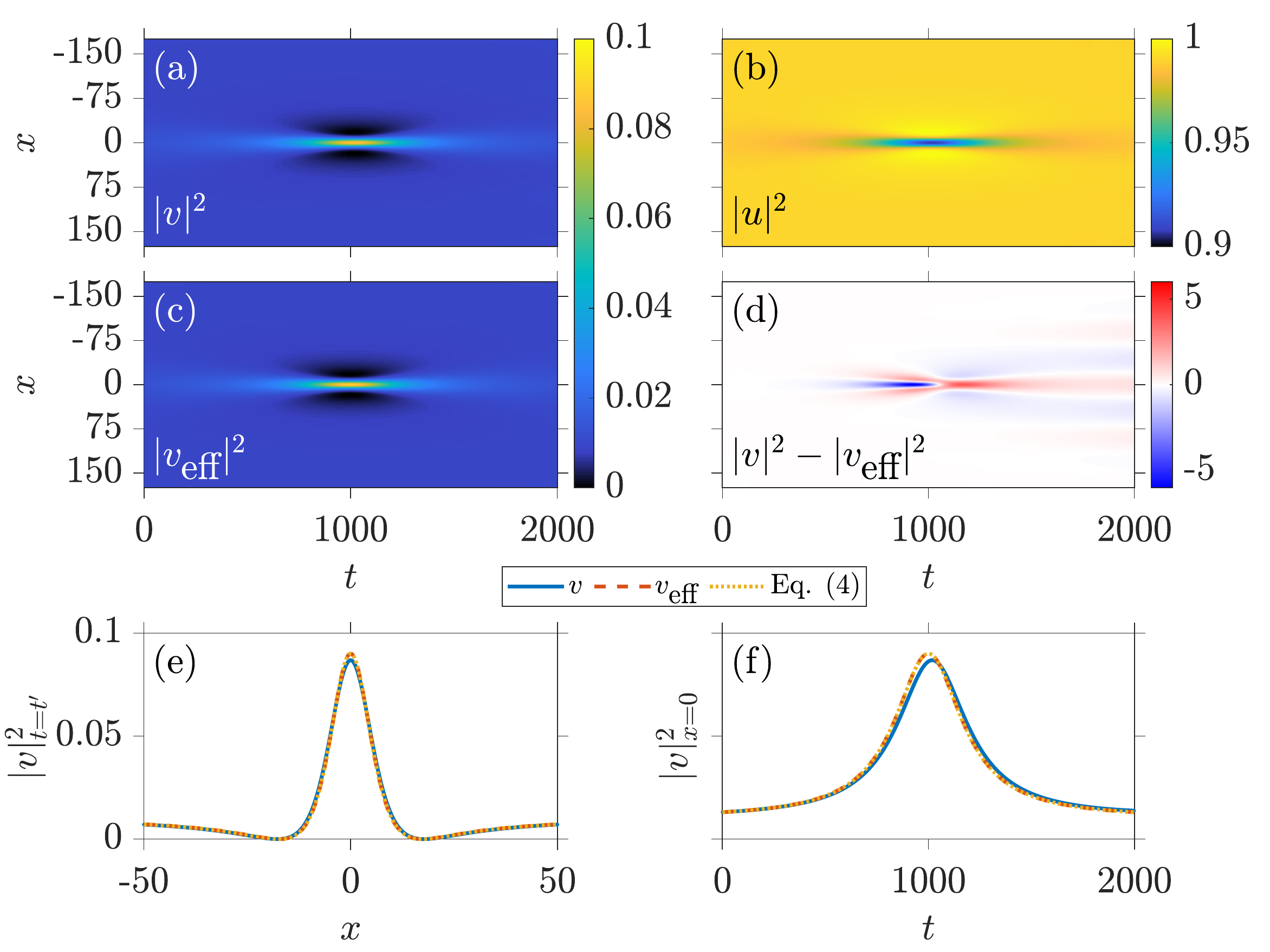}
    \caption{
    {Density evolution of the Peregrine initial condition [Eq.~\eqref{eq:peregrine}] in the absence of a trap for (a) the minority and (b) the majority species of the two-component setting with $\mu_u=1$.
    Equation \eqref{eq:peregrine} is initialized with $P_o=0.01$ and $t_o=1000$.
    (c) Dynamics of the density within the effective single-component model.
    (d) Density difference between the (a) two- and the (c) single-component dynamics (the colorbar is rescaled by a factor of $10^{3}$).
    (e) Density profile of the $v$ component for the two- and single-component setups at the time instant of Peregrine formation, namely $t_\textrm{two}'=1019$ and $t_\textrm{single}'=1000$.
    (f) Temporal evolution of the Peregrine wave emerging in the $v$ component of both the two- and single-component setups at $x=0$ (see legend).}
    In both panels (e) and (f), the corresponding analytical Peregrine solution of Eq.~\eqref{eq:peregrine} is provided.
    Note that length and time are measured in units of $a_\perp$ and $\omega_\perp^{-1}$, respectively.
    }
    \label{fig:peregrine_homo}
\end{figure}
\begin{figure}[t]
    \centering
    \includegraphics[width=\linewidth]{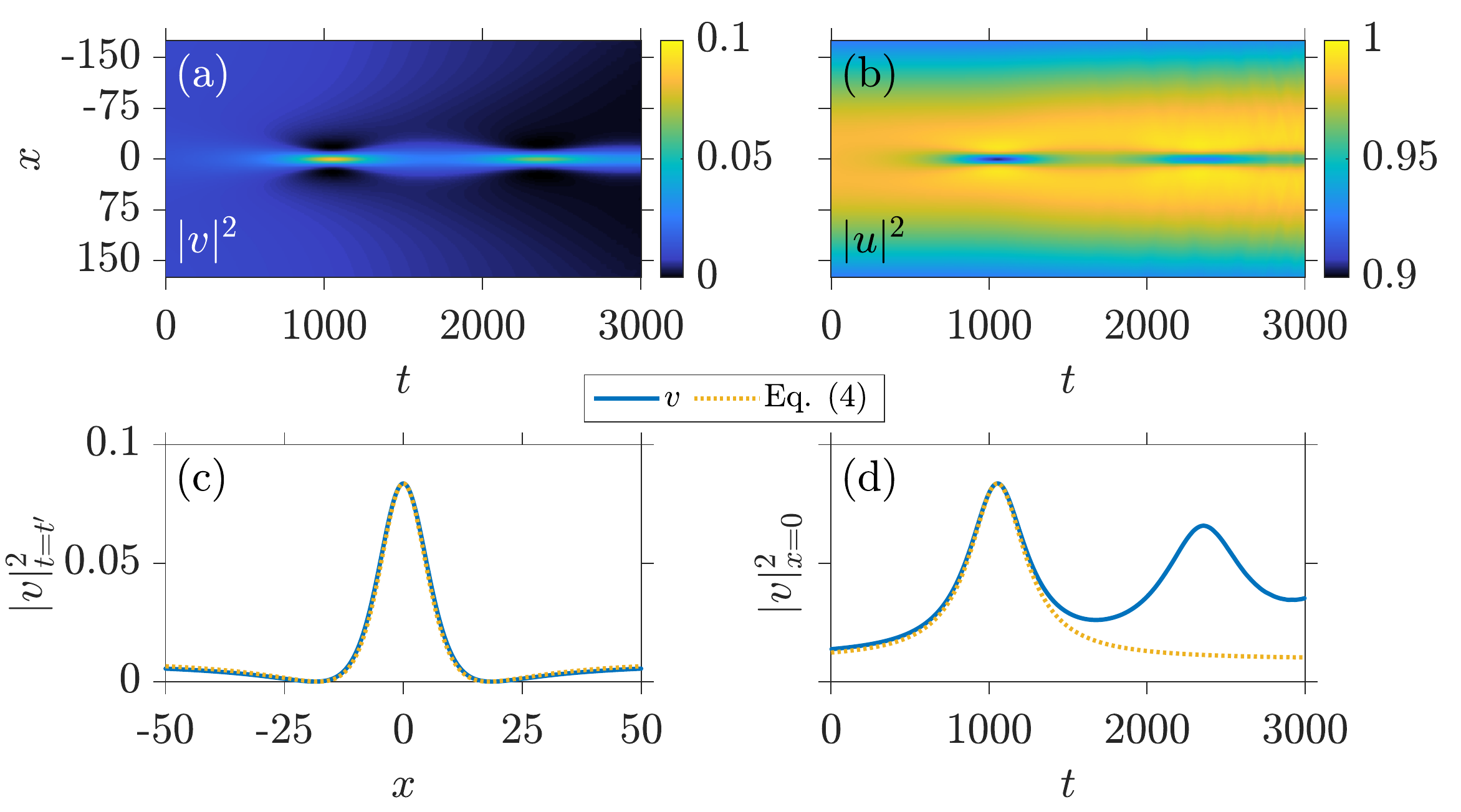}
    \caption{
    Spatiotemporal density evolution of a Peregrine initial condition 
    in the presence of a trap with $\Omega=0.002$ for (a) the minority and 
    (b) the majority species of the two-component case with $\mu_u=1$.
    Equation ~\eqref{eq:peregrine} is initialized with $P_o=0.01$ and $t_o=1000$ on top of the ground state obtained for the super Gaussian potential of Eq.~\eqref{eq:supergaussian} using $\mu_v=P_o$ (see text).
    (c) Density snapshot of the $v$ component capturing the instantaneous formation, $t_\textrm{two}'=1053$, of the Peregrine soliton.
    (d) The temporal density profile of the $v$ component at the fixed spatial location of $x=0$.
    In both panels (c) and (d), the relevant theoretical prediction of Eq.~\eqref{eq:peregrine} is provided for a direct comparison (see legend).
    Length (time) is given in terms of $a_\perp$ ($\omega_\perp^{-1}$).
    }
     \label{fig:peregrine_trap}
\end{figure}

\section{Dynamics of the Peregrine soliton} \label{sec:results}
\subsection{Proof-of-principle} \label{sec:proof-of-principle}

First, we consider the scenario of a Peregrine initial condition in the absence of a trap in both the two-component setup, described by Eqs.~\eqref{eq:CGPE_u} and \eqref{eq:CGPE_v} 
and within the effective single-component framework of Eq.~\eqref{eq:eff_CGPE}.
Here, we initialize the minority $v$ component and effective single-component one with Eq.~\eqref{eq:peregrine}, with $P_o=0.01$ and $t_o=1000$. 
In the two-component setup, the majority $u$-component wave function has the 
form $u(x)=\sqrt{\mu_u-|v(x)|^2}$, while the chemical potential $\mu_u=1$ is held fixed.
The relevant dynamics is shown in Fig.~\ref{fig:peregrine_homo}.
In terms of the minority $v$ component, the evolution of the two-component setup clearly follows that of the effective single-component one [cf. Figs.~\ref{fig:peregrine_homo}(a) and \ref{fig:peregrine_homo}(c), respectively].
Note also how the majority $u$ component in Fig.~\ref{fig:peregrine_homo}(b) naturally accompanies the density bump of the Peregrine formation with a corresponding (complementary) density dip.
The latter is reminiscent of dark-bright (DB) solitonic entities~\cite{Kevrekidis2016Solitons} and naturally stems from the repulsive interaction between the components. 

For a more accurate comparison of the two models, the density difference of the $v$ components is shown in Fig.~\ref{fig:peregrine_homo}(d).
Here, one can observe that around the Peregrine formation the two- and {the} single-component setups differ {at most by $|v|^2-|v_\textrm{eff}|^2 \sim 10^{-3}$ (recall that the color bar is rescaled by a factor of $10^3$).} 
In particular, their difference may grow, but this only happens past the initial formation of the Peregrine soliton. 
This is rather natural to expect, given that for large densities the correction to the single-component approximation of $v$ becomes substantial. 
As discussed in Ref.~\cite{Bakkali-Hassani2021}, such a correction is proportional to 
$\left(\sqrt{\mu_u-|v|^2}\right)_{xx}/\left(2\sqrt{\mu_u-|v|^2}\right)$ and can thus become relevant when the spatially varying wave form of $|v|^2$ grows
substantially, 
upon the emergence of the Peregrine.

Indeed, by inspecting in Fig.~\ref{fig:peregrine_homo}(e) each Peregrine soliton at the instant of its formation, $t'$, it is observed that the Peregrine of the two-component system is slightly smaller in amplitude and slightly wider when compared to the one formed in the single-component setting.
Also, as shown in Fig.~\ref{fig:peregrine_homo}(f), we find that the Peregrine formation in the two-component setup takes place later than in the single-component one.
In particular, the former occurs at $t'_\textrm{two}=1019$, while the latter occurs at $t'_\textrm{single}=1000$.
In Figs.~\ref{fig:peregrine_homo}(e) and \ref{fig:peregrine_homo}(f), we also provide the {analytical solution of} Eq.~\eqref{eq:peregrine}, which, as expected, falls on top of the {wave form stemming} from the single-component setup.
Hence, we can conclude that in the two-component scenario the presence of the majority component slightly hinders and/or delays the Peregrine formation. Additionally, a closer inspection
of Fig.~\ref{fig:peregrine_homo}(d) reveals progressively stronger deviations between the single- and two-component description past the Peregrine formation time.
This stems from the differences in the manifestation of the modulational instability of the background for large and positive times (see, for example, Ref.~\cite{Charalampidis2018a} for a discussion of the relevant instability in the presence of the Peregrine wave).
Moreover, we also found that at those large positive times a wedge-like structure emerges, similar to that described in Ref.~\cite{Biondini2016b} and, more recently, experimentally found in Ref.~\cite{Kraych2019} (see also Appendix~\ref{app:wedge-structure}).

In the presence of a parabolic trapping potential ($\Omega=0.002$) we prepare our initial state as follows.
On the one hand, we obtain the single-component ground state, $u_\textrm{GS}$, of the majority $u$ component for $\mu_u=1$.
On the other hand, we find the ground state, $v_\textrm{GS}$, of the minority $v$ component in a super Gaussian trapping potential of the form
\begin{align}
    V_\textrm{SG}(x) = 1-\exp[-\qty(\frac{x}{r_\textrm{TF}})^{100}]\,,
    \label{eq:supergaussian}
\end{align}
but for $\mu_v=P_o=0.01$, which throughout the text refers to the chemical potential of the minority $v$ component.
For this initially decoupled two-component system, $r_\textrm{TF}=\sqrt{2\mu_u}/\Omega$ in Eq.~\eqref{eq:supergaussian} denotes the Thomas-Fermi radius of the majority $u$ component.
From here, we imprint Eq.~\eqref{eq:peregrine} with $t_o=1000$ onto the minority ground state as $v(x)=v_\textrm{GS}(x)v_\textrm{eff}(x,0)/\sqrt{P_o}$.
The reason for using a super Gaussian is to obtain an initial state with an almost constant (flat) background {achieving also a smoother decay of the tails of the Peregrine,} as described by Eq.~\eqref{eq:peregrine}.
Lastly, we construct the majority $u$ component wave function by subtracting the minority $v$ component from the majority single-component ground state, i.e., $u(x)=\sqrt{|u_\textrm{GS}(x)|^2-|v(x)|^2}$.
Note that the latter operation emulates a particle transfer from the majority component to the minority one, while keeping the total density constant, which experimentally can be implemented by means of a two-photon Raman transition, as in the recent Townes soliton realization~\cite{Bakkali-Hassani2021}.
Having carried out these three steps (ground state under super Gaussian, Peregrine imprinting,
and formation of the complementary majority component), we are ready to perform our direct numerical simulations in the presence of a parabolic trap.
During the dynamics, the super Gaussian is turned off, interspecies interactions are switched on, and both components evolve under  the {influence of the} same harmonic trapping potential.

The in-trap {dynamical evolution of the two-component system} is presented in Fig.~\ref{fig:peregrine_trap}.
Specifically, Figs.~\ref{fig:peregrine_trap}(a) and \ref{fig:peregrine_trap}(b) illustrate 
the spatiotemporal evolution of the density of the $v$ and $u$ components, respectively, showcasing the complementary nature of the latter. 
The spontaneous emergence of a Peregrine soliton, and of the corresponding density-dip appearing in the majority  $u$ component, takes place at $t'=1053$.
This precise time instant is captured in Fig.~\ref{fig:peregrine_trap}(c), where we further compare the emergent wave against the analytical solution of Eq.~\eqref{eq:peregrine}. 
The latter is fitted so as to match the maximum amplitude of the nucleated Peregrine, i.e., $P_o=|v_\textrm{max}|^2/9=0.0093$.
Notice that this value is rather proximal to the initial amplitude of the Peregrine wave form, namely $P_o=0.01$.
A nearly excellent agreement is observed between our numerical observation and the analytical estimate, an outcome that can also be inferred by inspecting Fig.~\ref{fig:peregrine_trap}(d).
In the latter, the evolution of the density of the $v$ component at $x=0$ is also depicted along with the theoretical prediction.
Indeed, the observed structure almost coincides around its core with the theoretically predicted one, but the numerically obtained one is found to be (very) slightly wider.
Similarly, small deviations between the two occur also in the far field (at the tails of the wave) but with their density difference never being greater than $\sim 10^{-3}$.
Even more importantly, at later evolution times the numerical solution recurs as a rogue pattern reminiscent of the Kuznetsov-Ma soliton~\cite{Kuznetsov1977,Ma1979,Kibler2012}.
Recall that the latter is a time-periodic family of solutions, of which the Peregrine is the asymptotic limit when the temporal periodicity tends to $\infty$.
A clear example of such a revival, that is related to the trapped setting at hand, can be seen in Fig.~\ref{fig:peregrine_trap}(a), but also in Fig.~\ref{fig:peregrine_trap}(d) around $t=2358$.

\begin{figure}[t]
    \centering
    \includegraphics[width=\linewidth]{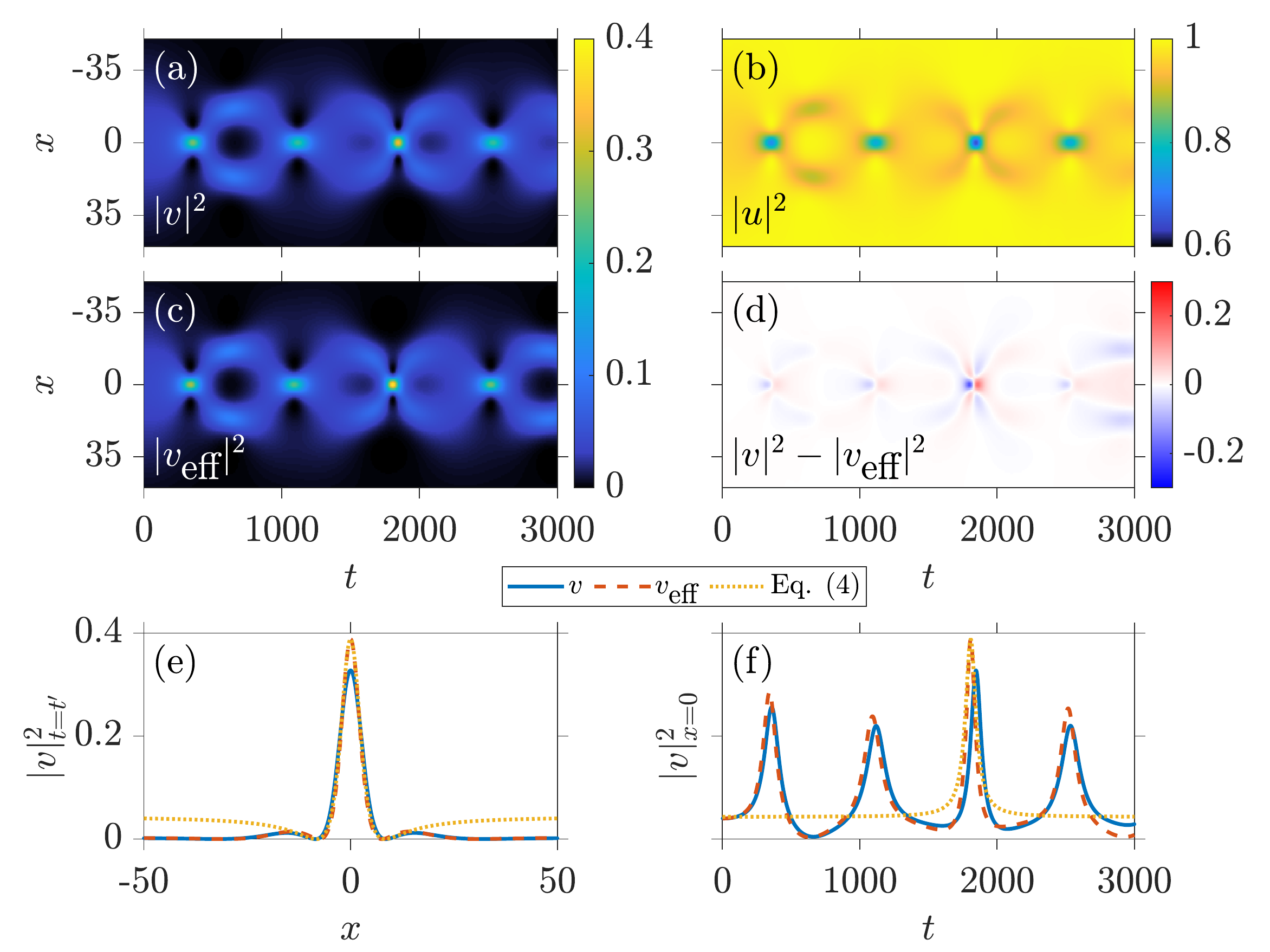}
    \caption{
    Time-evolution of the density of a Gaussian wavepacket with $A=0.2$ and $w=50$ depicting the dynamics of the (a) minority and (b) majority species of the two-component system with $\mu_u=1$.
    (c) Evolution of the minority $v$ component within the effective single-component description.
    (d) Density difference between the (a) two- and the (c) single-component dynamics. 
    (e) Density profile of the $v$ component for the two- and single-component setups at the times of their largest amplitudes, namely for $t'_\textrm{two}=1846$ and $t'_\textrm{single}=1808$, respectively. 
    (f) Evolution of the center position $x=0$ of the density of the $v$ component for the two- and the single-component setups.
    In both panels (e) and (f), the corresponding analytical solution of Eq.~\eqref{eq:peregrine} is given, with its peak fitted to the single-component case (see text).
	Length and time are expressed in units of $a_\perp$ and $\omega_\perp^{-1}$, respectively.
	}
    \label{fig:gauss_50}
\end{figure}
\begin{figure}[t]
    \centering
    \includegraphics[width=\linewidth]{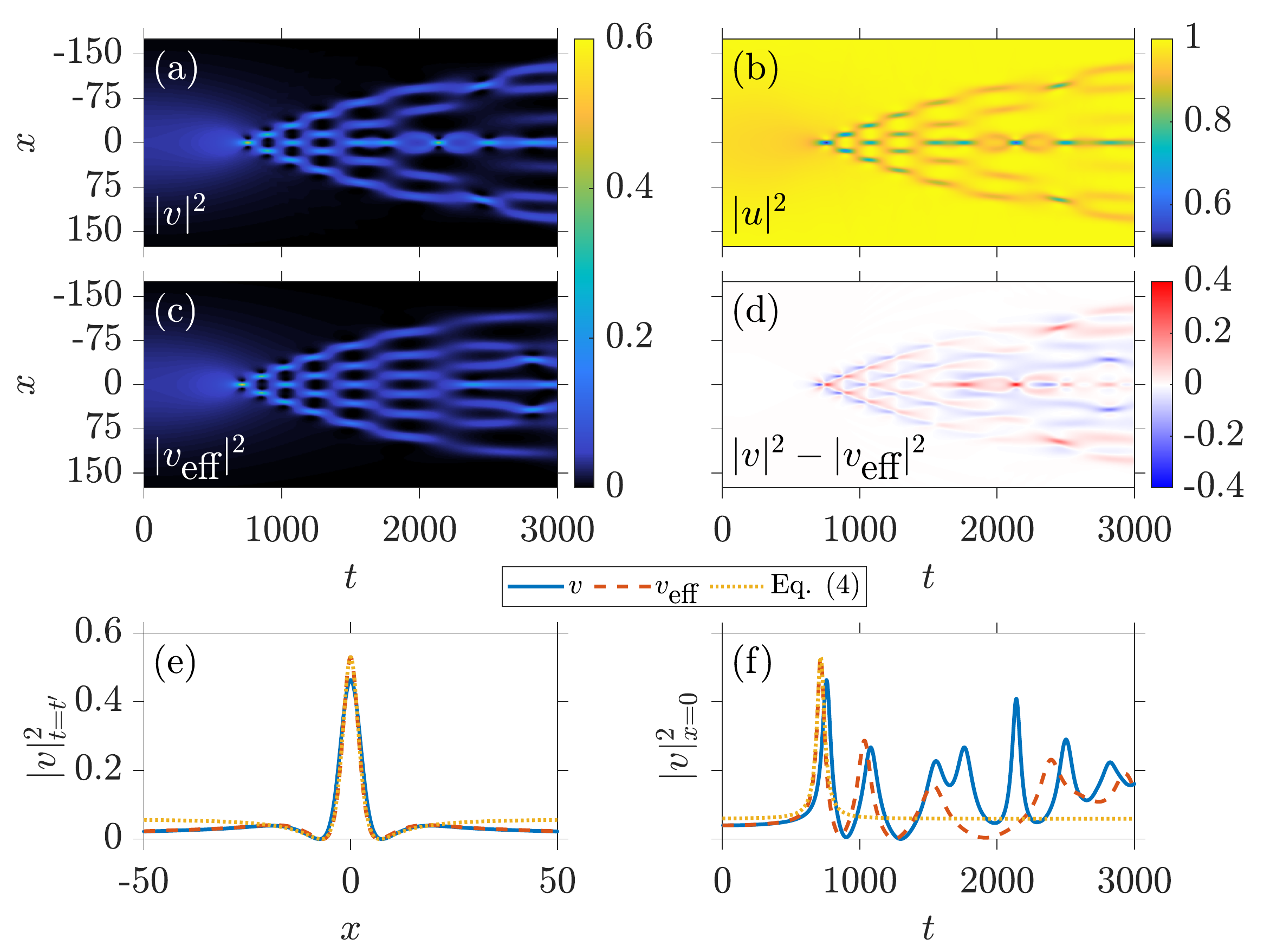}
    \caption{
   	Same as Fig.~\ref{fig:gauss_50} but considering a significantly broader Gaussian initial condition with $w=150$.
   	In this case, the Peregine soliton formation within the two models occurs at $t'_\textrm{two}=758$ and $t'_\textrm{single}=716$, respectively.
    Notice also the subsequent emergence of the Christmas-tree structure.
    Recall that length and time are given in units of $a_\perp$ and $\omega_\perp^{-1}$, respectively.
    }
    \label{fig:gauss_150}
\end{figure}

\subsection{Semiclassical regime: Gaussian profile} \label{sec:gaussian}

We now turn to the example of a Gaussian initial condition in order to cement the generic nature of the Peregrine soliton formation, as well as to showcase an example of an initial condition that could be (far more) straightforwardly accessible in BEC experiments.
The Gaussian profile here is representative of a wide initial condition in the minority $v$ component, so as to capture the semiclassical limit of the work of Ref.~\cite{Bertola2013}. As in the preceding section, we initialize the dynamics of the two- and effective single-component system under consideration first in the absence and subsequently in the presence of a harmonic confinement. 
Additionally, two different representative case examples corresponding to two distinct widths, namely $w=50$ and $w=150$, of the initial Gaussian wavepacket are considered, while in both cases the amplitude, $A=0.2$, of this Gaussian initial condition is held fixed.

The first case example, i.e., that of a narrower Gaussian initial condition ($w=50$), is shown in Fig.~\ref{fig:gauss_50}.
Here, the spatio-temporal evolution of the $v$ component of both the two- and the single-component setups [see Figs.~\ref{fig:gauss_50}(a) and \ref{fig:gauss_50}(c), respectively] present the same \textit{chain}like temporal pattern within which periodic recurrences of a localized pattern reminiscent of the Peregrine wave take place.
Also, the majority $u$ component of the two-component model, illustrated in Fig.~\ref{fig:gauss_50}(b), showcases the complementary \textit{dark chain} temporal pattern.
The array of individual rogue-wave patterns can be clearly discerned and once again there is a close correspondence between the single- and two-component dynamics.
To demonstrate this correspondence, the density difference of the $v$ components is provided in Fig.~\ref{fig:gauss_50}(d).
Clearly, the larger deviation between the two systems occurs around the location of the formation of the localized peaks ($x=0$).
To further expand on the comparison of the observed  structures, in Fig.~\ref{fig:gauss_50}(e) we illustrate their density profiles at the formation of their maximum, namely at $t'_\textrm{two}=1846$ and $t'_\textrm{single}=1808$, respectively, as compared with an exact Peregrine solution.
In particular, the largest among the localized patterns that emerged during the evolution is considered here. 
Notice that an adequate agreement is observed in the vicinity of the core, with the two-component pattern being slightly wider but also having a smaller amplitude when compared to the  single-component Peregrine wave that follows the theoretical prediction [see Eq.~\eqref{eq:peregrine}].
The latter, is fitted to the Peregrine of the single-component setup, having $P_o=|v_\textrm{eff}^\textrm{max}|^2/9=0.043 \approx A^2$ and selecting $t_o=t'_\textrm{single}$. 
Naturally, the Gaussian evolution, given its decaying tail, cannot lead to the constant background of the Peregrine wave form, hence the observed deviations in the far field.
Additionally, monitoring in Fig.~\ref{fig:gauss_50}(f) the density at $x=0$ in the course of the evolution shown in Figs.~\ref{fig:gauss_50}(a) and \ref{fig:gauss_50}(c) reveals that the emergent recurring structures in the chain slightly differ in their time of formation. 
Namely, within the effective model, each member of the chain (i.e., each recurrence event) appears earlier in time, presenting also a larger amplitude spike, when compared to the two-component setting.
The aforementioned results are in line with the trends of the ones found in the homogeneous case [see Figs.~\ref{fig:peregrine_homo}(e) and \ref{fig:peregrine_homo}(f)].

As a second representative example, a significantly wider, with $w=150$, Gaussian initial condition is considered again both in the two- and in the effective single-component setups (see Fig.~\ref{fig:gauss_150}).
It turns out that increasing the width of the Gaussian leads to the dynamical formation of the Christmas-tree structure (see Sec.~\ref{sec:setup}).
Indeed, as also explained previously, 
the mechanism underlying the fragmentation of the solution is the existence of strong modulational instability of the (in our case, effectively) focusing NLS equation~\cite{Bertola2013}.
The various asymptotic regions (smooth versus. highly oscillatory) are separated by ``breaking curves", and the solution of the focusing NLS inside the oscillation region is approximately given by modulated genus-2 waves. 
More specifically, as shown in Ref.~\cite{Bertola2013}, near each oscillation peak the solution takes on (locally) the universal shape of the rational solution known as the Peregrine soliton.
In our computations, a Christmas-tree pattern is clearly seen in Figs.~\ref{fig:gauss_150}(a) and \ref{fig:gauss_150}(c), corresponding to the two- and single-component setups, respectively,
while the complementary \textit{dark} Christmas-tree structure appearing in the majority $u$ component of the two-component system is presented in Fig.~\ref{fig:gauss_150}(b).
Notice here the ramifications of the emerging Christmas-tree structure right after the formation of the Peregrine soliton at $t'_\textrm{two}=758$ ($t'_\textrm{single}=716$) within the two-component (single-component) model.

The relevant Peregrine profiles are depicted in Fig.~\ref{fig:gauss_150}(e) at the time instant of their formation together with the analytical Peregrine solution of Eq.~(\ref{eq:peregrine}).
Recall that the latter is fitted so as to match the amplitude of the single-component Peregrine wave.
Also in this case an excellent agreement is observed around the waves' core when comparing the exact solution to the single- and the two-component outcome.
Thus, increasing the width of the initial Gaussian profile leads to an overall increase in size of the emerging Peregrine soliton and to the formation of the Christmas-tree structure.
However, so as to stretch the comparison of the patterns appearing in the distinct settings, in Fig.~\ref{fig:gauss_150}(f) we illustrate the temporal evolution of the central density of both $v$ components. 
Evidently, this quantity perfectly captures the instant where the dynamics begins to differ,
i.e., $t\sim1500$ [cf. Figs.~\ref{fig:gauss_150}(a) and \ref{fig:gauss_150}(c)].
The observed discrepancy between the two models is a direct consequence of the presence of the majority $u$ component in the two-component setup. 
This is an effect which will be even more pronounced in the case of the presence of harmonic confinement that follows.

Next, we extend the above Gaussian state considerations to the case where a parabolic trap, with trapping frequency $\Omega=0.002$, is also present.
Also in this case, the initial state preparation consists of obtaining the decoupled, single-component ground state of the majority $u$ component for a fixed chemical potential ($\mu_u=1$), and then approximating the particle transfer to the minority $v$ component by subtracting the latter from the former.
It turns out that the dynamical evolution of narrower Gaussian wave packets leads to qualitatively identical results to those found in the relevant homogeneous investigations discussed above, and thus these findings are not included herein for brevity.
\begin{figure}[t!]
    \centering
    \includegraphics[width=\linewidth]{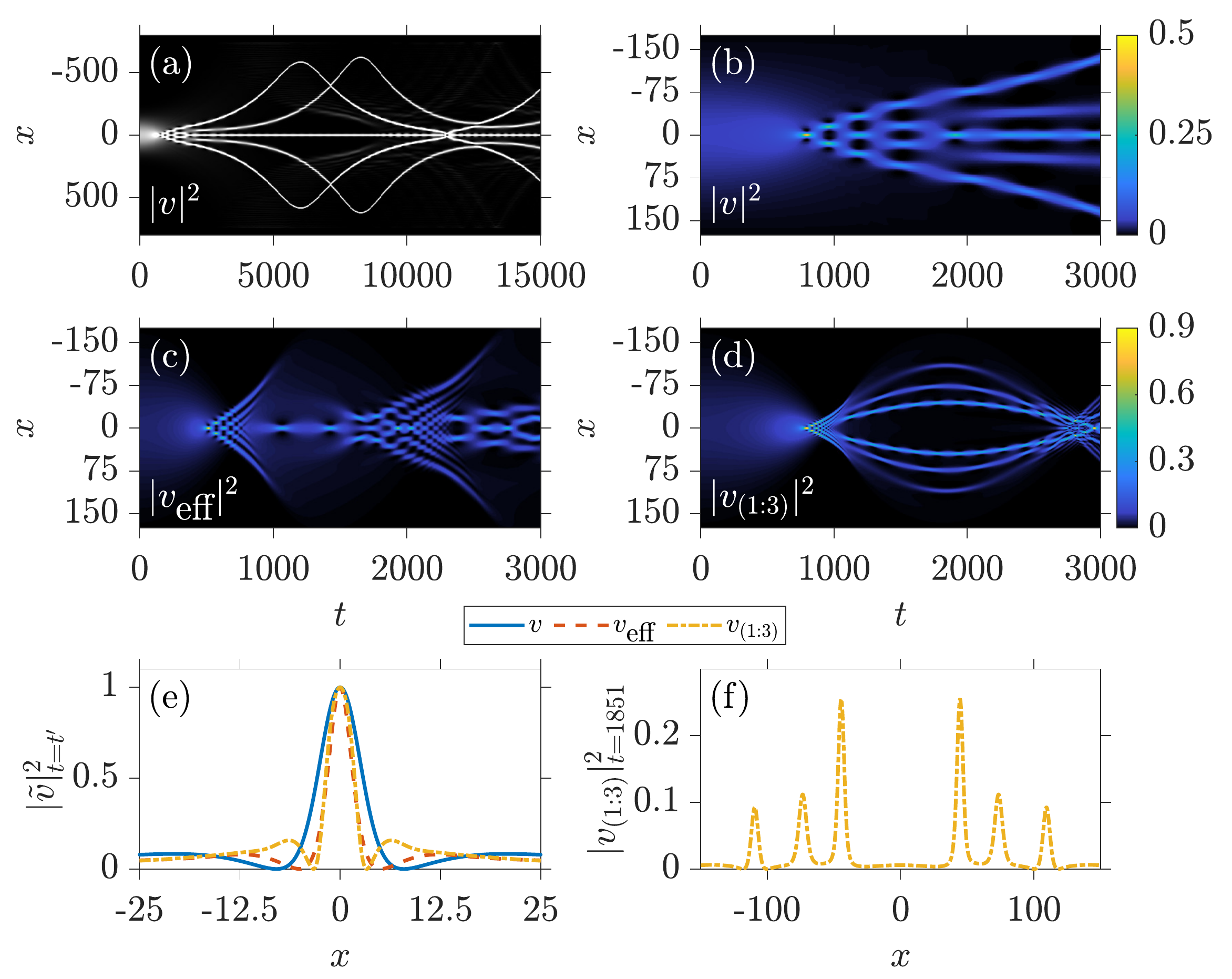}
    \caption{
    Evolution of the density of a Gaussian wavepacket with $A=0.2$ and $w=150$ in the presence of a harmonic trap with $\Omega=0.002$. 
    (a) Temporal evolution of the minority $v$ component. 
    (b) A magnification of (a) is provided.
    (c) Effective single-component case.
    (d) Same as (b) but upon considering a mass-imbalanced BEC mixture (see text). 
    (e) Density profiles of the $v$ component for the single-component, and mass-balanced and mass-imbalanced two-component setups at the instant of 
    formation of the Peregrine soliton $t'_\textrm{single}=516$, $t'_\textrm{two}=800$, and $t'^{(1:3)}_\textrm{two}=795$, respectively.
    All Peregrine amplitudes are normalized to unity for direct comparison.
    (f) Density profile at $t'^{(1:3)}_\textrm{two}=1851$, depicting the pairwise DB soliton nucleation 
    that follows the decay of the Christmas-tree pattern,
    for the mass-imbalanced mixture.
    In all cases, the majority $u$ component complements its relevant minority one, and thus it is omitted.
    The length and time units shown are in terms of $a_\perp$ and $\omega_\perp^{-1}$, respectively.}
    \label{fig:gauss_150_trap_MI}
\end{figure}

On the contrary, a more complex evolution takes place when a wider ($w=150$) Gaussian initial condition is considered. 
Here, we observe an interesting phenomenon that we did not encounter in the previous settings, so it is convenient for our purposes to monitor the dynamics both at long [Fig.~\ref{fig:gauss_150_trap_MI}(a)] and short timescales [Fig.~\ref{fig:gauss_150_trap_MI}(b)]. 
At initial times, and in particular around $t'_\textrm{two}=800$, the Peregrine soliton forms, being subsequently followed by the emergence of the Christmas-tree structure.
Note that the Peregrine formation takes place at later times as compared to the relevant untrapped scenario. 
Also, in this case the presence of a modulated density profile does not appear to sustain this Christmas-tree structure, which is seen to break into several DB soliton-like entities after $t=2000$. 
Following these soliton-like structures for longer evolution times [Fig.~\ref{fig:gauss_150_trap_MI}(a)] reveals that these patterns oscillate inside the parabolic trap over a \textit{very} large period. In particular, their oscillation frequency is substantially smaller than that of the trap.
Also, their corresponding period is much longer than that expected for regular DB solitary waves under the same confinement conditions~\cite{Busch2001,Kevrekidis2015}, which would typically be of the order of a few hundred time units.
By comparing the trajectories of these peculiar DB entities to regular ones, we observe a distinct concavity between the two. 
Namely, the present patterns feature (unprecedented, to our knowledge) convex trajectories until very close to the turning point, contrary to the usual (nearly) harmonic oscillations experienced by standard DB solitons.
Nevertheless, after an extremely long period of time and around $t=11394$ the two outermost individual patterns interfere to produce a revival of a Peregine wave form. While the latter is not identical to the early one formed at $t'_\textrm{two}=800$, it is very proximal to 
a Peregrine pattern having a distinct amplitude.
Hence, the system can access a very long time scale revival of the relevant pattern in this parabolically trapped setting that is particularly interesting 
in its own right. 
Recall also that in all of the above cases the majority $u$ component evolution complements the minority $v$ component shown in Figs.~\ref{fig:gauss_150_trap_MI}(a) and \ref{fig:gauss_150_trap_MI}(b).
Hence, it is not included here.

For completeness, the effective single-component evolution is illustrated in Fig.~\ref{fig:gauss_150_trap_MI}(c).
Here, the dynamics appears to be dramatically faster when compared to the aforementioned two-component scenario, with the Peregrine soliton appearing at $t'_\textrm{single}=516$ and being of higher amplitude [cf. Figs.~\ref{fig:gauss_150_trap_MI}(b) and Fig.~\ref{fig:gauss_150_trap_MI}(c) color bars on the right side].
Also, faster is the formation of the corresponding Christmas-tree structure which once more, due to the presence of the trap, cannot be sustained and around $t\sim1000$ it blurs out into a smooth background within which the recurrence of two Peregrine solitons, one at $t=1068$ and one at $t=1425$, is evident.
Note that the smoothing of the Christmas-tree pattern is a unique feature of this effective single-component setup.
Strikingly enough, also a recurrence of a second larger Christmas-tree structure is observed in this single-component setting, around $t\sim2500$, after an interval where interference processes take place.
This, in turn, confirms the fact that in the two-component setting, the majority $u$ component plays a major role, not only on the speed of the events, but also in the formation of more complex and robust structures, such as the numerically observed DB soliton-like waves found in this work.

\subsection{Impact of the mass-imbalance}

As a next step, our aim is to generalize our findings by considering mixtures in which the two species bear different masses. 
In particular, in the way of a concrete example, here we  focus our investigations on mass-imbalanced mixtures having a mass ratio of (1:3) to gain a qualitative overview of the main phenomenology, while mimicking the potentially experimentally relevant situation of heteronuclear BEC mixtures of, e.g., $^{87}$Rb--$^{174}$Yb, $^{23}$Na--$^{7}$Li, or $^{87}$Rb--$^{23}$Na atoms. 
Although the initial state preparation considered in this case is the same as before, it is important to note that the trapping potential will now affect each component differently [see Eq.~\eqref{eq:CGPE}]. 
In general, it is found that if the majority component is the heaviest one, Peregrine wave generation, similar to that shown in Fig.~\ref{fig:peregrine_trap}, occurs only for narrow initial Gaussian wave functions ($w \lesssim 60$). Therefore, the Christmas-tree structure is absent and in particular for $w > 60$ a delocalization of the Gaussian takes place.
On the other hand, if the minority component is the heaviest one we retrieve the overall phenomenology, namely that of the Peregrine soliton formation being followed by the nucleation of a Christmas-tree pattern. 
Interestingly, the value of the Gaussian width above which the Christmas-tree pattern appears is affected by the inter-component mass ratio.
For instance, for  $ m_u/m_v=1/3$ it occurs for $w>30$, while in the mass-balanced case occurs for $w>52$.

A case example is shown in  Fig.~\ref{fig:gauss_150_trap_MI}(d) for an initial Gaussian profile with $A=0.2$ and $w=150$.
For mass-imabalanced mixtures, the dynamical rogue-wave pattern formation ($t'=795$) 
is slightly accelerated when compared to the equal mass scenario discussed above [cf. Fig.~\ref{fig:gauss_150_trap_MI}(a)].
Moreover, as mentioned earlier, the heaviest minority component experiences now a tighter trapping potential. 
This has as a result the emergence of a Peregrine wave that has its side humps  higher 
in amplitude and closer to the core of the structure. 
A direct comparison of the Peregrine soliton formed in this setting with the relevant waves generated within the mass-balanced and the effective single-component models is provided in Fig.~\ref{fig:gauss_150_trap_MI}(e).
Furthermore, after the decay of the Christmas-tree structure, we again observe DB soliton-like structures being emitted around $t\sim 1000$.
{The first half oscillation period of these configurations can be seen in Fig.~\ref{fig:gauss_150_trap_MI}(d), and the corresponding bright solitary-wave density profile is} depicted in  Fig.~\ref{fig:gauss_150_trap_MI}(f) at $t=1851$, i.e., at the maximum amplitude of its oscillation. 
Besides their oscillation inside the parabolic trap, these structures further undergo a periodic amplitude breathing. 
The frequency of this breathing is found to increase with the size of the soliton; i.e., it appears to be larger for the solitary-waves that are closer to the trap center.
Additionally,
it is also found that these structures emerge always in counterpropagating pairs, the number of which is directly proportional to the width of the initial Gaussian profile. 
For instance, for $w=50$ ($w=150$) the number of counterpropagating pairs is one (three).

\begin{figure}[t!]
    \centering
    \includegraphics[width=\linewidth]{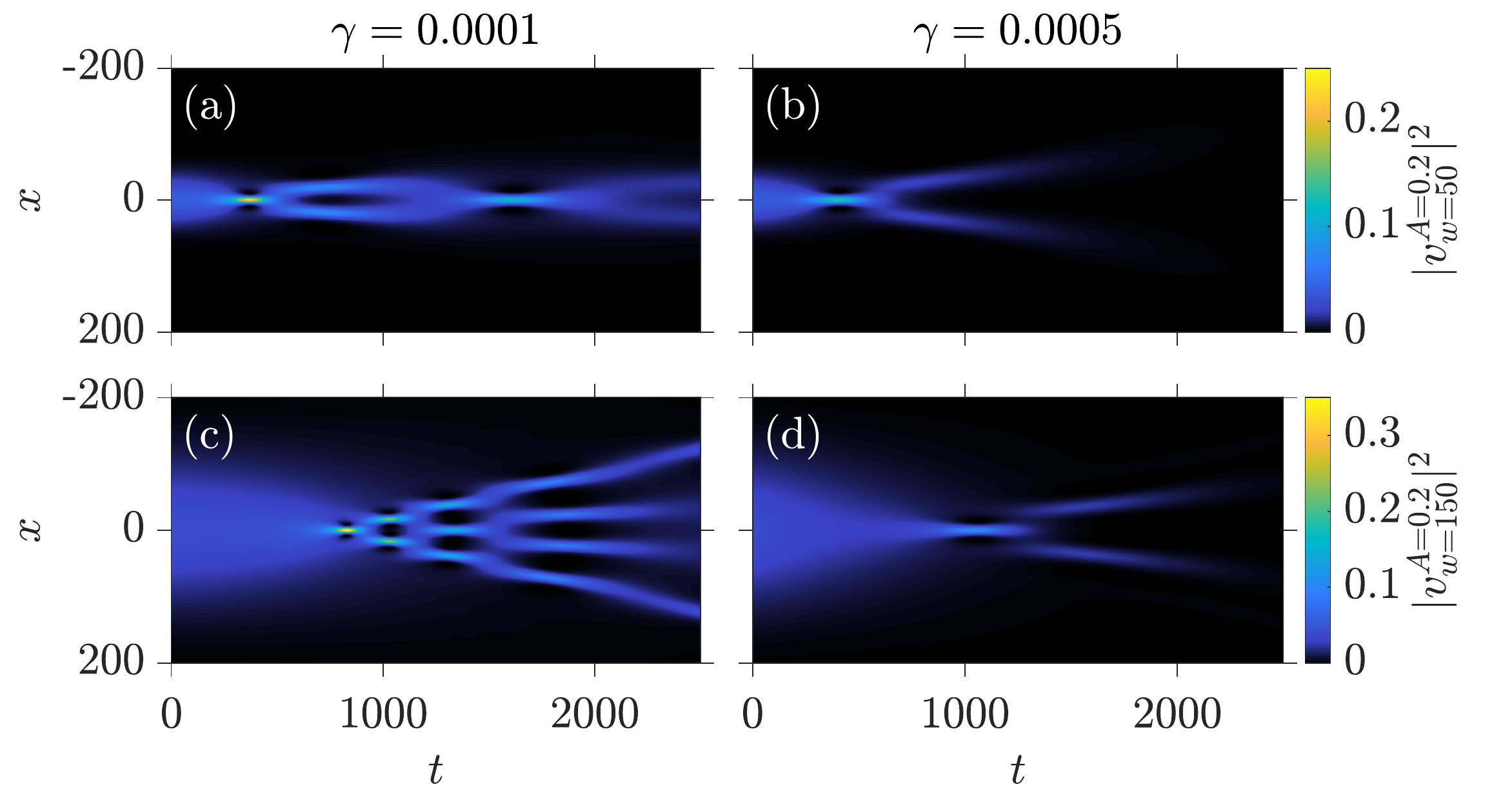}
    \caption{ Dynamical evolution of the density of the minority $v$ component for $A=0.2$ and [(a), (b)] $w=50$ and [(c), (d)] $w=150$. The dissipation strength is [(a), (c)] $\gamma=10^{-4}$ and [(b), (d)] $\gamma=5 \times 10^{-4}$. 
    Other parameters used are $\mu_u=1$, $\mu_v=0.01$, $m_u=m_v$, and $\Omega=0.002$. 
    Note that length (time) is in units of $a_\perp$ ($\omega_\perp^{-1}$).}
    \label{fig:dissipation}
\end{figure}
\begin{figure*}[t!]
\centering
\includegraphics[width=\linewidth]{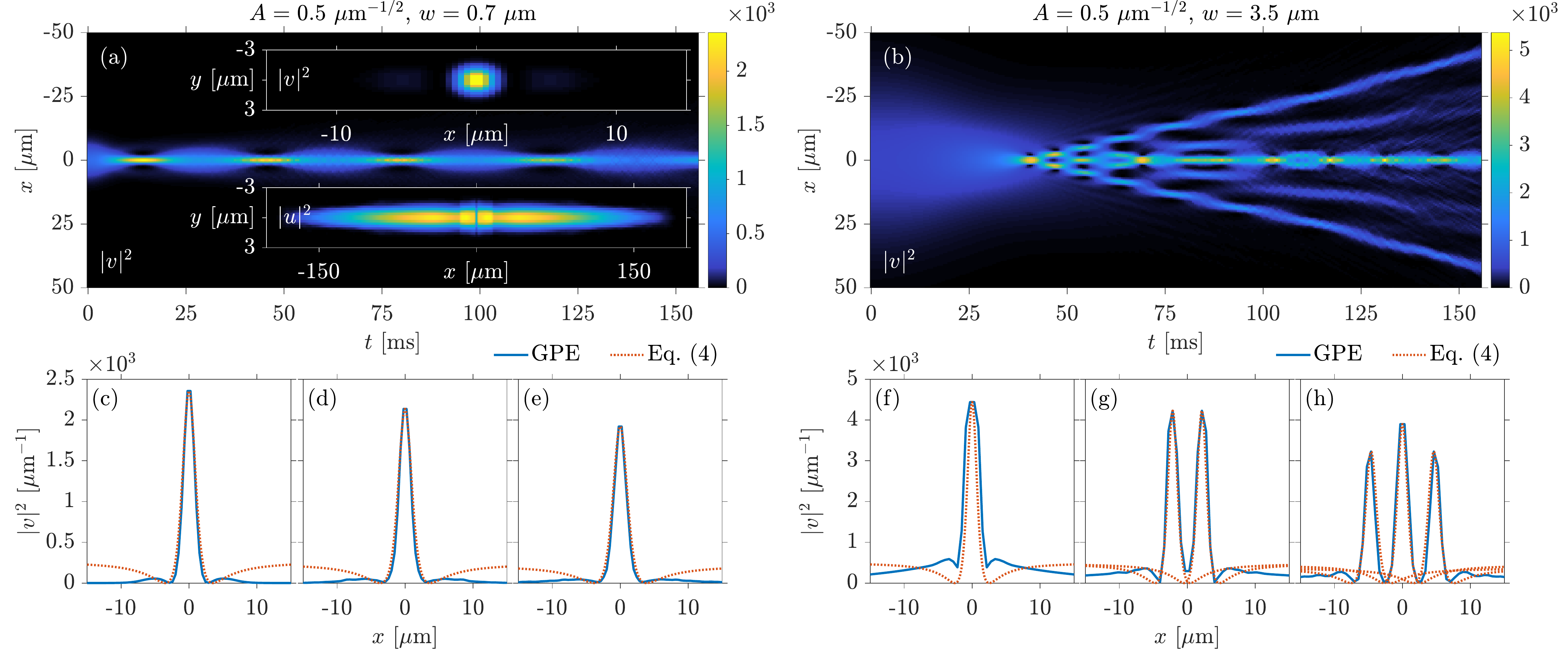}
\caption{ 
	(a) [(b)] Spatiotemporal evolution of the integrated along the $yz$-directions density, of the minority $v$ component for $A=0.5~{\rm \mu m^{-1/2}}$ and $w=0.7~{\rm \mu m}$ [$w=3.5~{\rm \mu m}$]. 
	Insets in panel (a) depict density snapshots of the (top) $v$- and (bottom) $u$ components, rescaled by a factor of $1/3$ and $1/7$, respectively, in the $x$-$y$ plane at $t_0 \approx 13.5\,$ms when the Peregrine initially forms.
	Panels (c)--(e) [(f)--(h)]: Integrated density profiles at selective time instants during evolution of the Peregrine and its revivals [of the Christmas tree].
	A comparison with the relevant analytical prediction of Eq.~(\ref{eq:peregrine}) is also provided (see legends).    
	For this quasi-1D evolution, the corresponding trapping frequencies are $(\omega_x, \omega_y, \omega_z)=2\pi \times (3.06, 400, 400)\,$Hz and $N=7 \times 10^5$.}
\label{fig:3d}
\end{figure*}

\section{Thermal effects on Peregrine generation}\label{imperfections_dyn}

Having established Peregrine soliton generation 
stemming from generic initial conditions, below, 
we shall exemplify the validity of our findings in the 
presence of dissipation. To this end, we consider 
the general system of two coupled dissipative GPEs~\cite{Proukakis2008,Yan2014Exploring,Cockburn2010MatterWave} for the mass-balanced mixture 
\begin{subequations}
\label{eq:DCGPE}
\begin{align}
\left (i - \gamma\right)u_t = -\frac{1}{2} u_{xx} + V_u(x) u + (g_{11} |u|^2 + g_{12} |v|^2) u \,,
\label{eq:DCGPE_u}
\\
\left(i - \gamma \right)v_t = -\frac{1}{2} v_{xx} + V_v(x) v + (g_{21} |u|^2 + g_{22} |v|^2) v \,,
\label{eq:DCGPE_v}
\end{align}
\end{subequations}
In these equations, $\gamma$ represents the  
dimensionless dissipation strength which is in turn 
related to the system’s temperature. 
For instance, when $\gamma \ll 1$ lies in the interval 
[$2 \times 10^{-4}$, $2 \times 10^{-3}$], it corresponds in dimensional units to temperatures 
$[10,100]\,$nK~\cite{Yan2014Exploring,Cockburn2009stochastic,Cockburn2010MatterWave}. 
The reduction of the above-mentioned system of dissipative GPEs follows the assumption that only
the thermal modes along the longitudinal direction are populated. 
The relevant dynamics of a trapped ($\Omega=0.002$) narrow Gaussian wavepacket 
having $A=0.2$, $w=50$ in the case of $\gamma=10^{-4}$ and $\gamma=5 \times 10^{-4}$ is presented 
in Figs.~\ref{fig:dissipation}(a) and \ref{fig:dissipation}(b), respectively. 
As previously, we only show the time evolution of the minority $v$ component density. 
The majority $u$ component is complementary to it while being significantly less affected by the presence of dissipation, at least within the considered timescales. 
Notice that even though the Peregrine soliton emerges around $t=400$ in both scenarios 
it is of smaller amplitude as $\gamma$ increases. 
Moreover, due to faster particle loss for larger 
$\gamma$'s, the pattern observed for smaller $\gamma$ 
is lost [see Fig.~\ref{fig:dissipation}(b)]. 
Dissipation affects in a similar manner also the 
nucleation of the Christmas-tree configuration.
The latter is generated only for 
$\gamma < 5 \times 10^{-4}$ 
when broader ($w=150$) Gaussian wave packets are used [Fig.~\ref{fig:dissipation}(c)], while its signature is 
lost for larger values of the dissipative parameter [Fig.~\ref{fig:dissipation}(d)]. 
In this latter situation only a single Peregrine occurs around $t \approx 1100$, i.e., at later times when compared to its faster (around $t \approx 750$) nucleation for $\gamma=10^{-4}$. In conclusion, nucleation of Peregrine solitons in repulsive media takes place for $\gamma<10^{-3}$ corresponding to temperatures smaller than $100\,$nK. 
However, more composite structures such as the above-discussed Christmas-tree and other observed (e.g., breathing) patterns are less robust, surviving only if $\gamma<5 \times 10^{-4}$, namely for temperatures roughly above $10\,$nK.

\section{Quasi-{\bf 1D} Peregrine and Christmas tree formation}\label{quasi1D_dyn}

Next, we aim to also testify Peregrine soliton nucleation in a  3D (yet {quasi}-1D) environment that can be readily implemented in recent experimental setups~\cite{Lannig2020Collisions,Bersano2018ThreeComponent,Katsimiga2020}. 
In this case, the dissipative effect stemming from the transverse directions on this intrinsically 1D wave will be appreciated. 
Specifically, we consider a system of 3D coupled GPEs:
\begin{subequations}
\label{eq:3DCGPE}
    \begin{align}
    iu_t = -\frac{1}{2} \nabla^2 u + V({\bf r}) u + (4 \pi a_{11} N_1 |u|^2 + 4 \pi a_{12} N_2 |v|^2) u \,,
    \label{eq:3DCGPE_u}
    \\
    iv_t = -\frac{1}{2} \nabla^2 v + V({\bf r}) v + (4 \pi a_{21} N_1 |u|^2 + 4 \pi a_{22} N_2 |v|^2) v \,,
    \label{eq:3DCGPE_v}
    \end{align}
\end{subequations}
that describes a mass-balanced binary mixture. 
The above set of equations is cast in this dimensionless form by rescaling space and time coordinates as  
$x'=a^{-1}_{\rm {ho}}x$, $y'=a^{-1}_{\rm {ho}}y$, $z'=a^{-1}_{\rm {ho}}z$, with $a_{\rm {ho}}=\sqrt{\hbar/m\omega_x}$ being the harmonic oscillator 
length along the longitudinal $x$ direction, and $t'=\omega_x t$.  
The corresponding wave functions are also rescaled as $u(x',y',z')=\sqrt{N_i/a^3_{\rm {ho}}}u(x,y,z)$ and 
$v(x',y',z')=\sqrt{N_i/a^3_{\rm {ho}}}v(x,y,z)$ where $N_i$ is the particle number per component. 
$\nabla^2 \equiv \left(\partial^2_x + \partial^2_y + \partial^2_z \right)$ is the Laplacian operator and $a_{jk}$ is the 3D scattering length accounting for the intra- ($j=k$) and inter-component ($j\neq k$) interactions.
In particular, the experimentally relevant $a_{11}=100.04a_0$, $a_{22}=95.44a_0$, and $a_{12}=a_{21}=110.0a_0$ are utilized with $a_0$ denoting the Bohr radius.
Additionally, $V({\bf r})=\frac{1}{2}\left( x^2+\left(\omega_y/\omega_x\right)^2 y^2+\left(\omega_z/\omega_x\right)^2 z^2 \right)$ is the 3D parabolic potential, where ${\bf r}=(x,y,z)$, with axial and transverse trapping frequencies assumed here to be $\left(\omega_x,\omega_y,\omega_z\right)=2\pi \times \left(3.06, 400, 400\right)\,$Hz. 
Notice that the above choice provides an aspect ratio $\omega_x/\omega_{\perp}\approx 0.008$ 
leading to a highly elongated (cigar-shaped) trapping geometry.

To initiate the dynamics in this 3D, yet quasi-1D setting we 
assume that the system contains $N=7 \times 10^5$ 
particles and that it remains in the ground state along the transverse $yz$-directions. 
Accordingly, each component's wave function at $t=0$ 
is expressed as: $u(x,y,z)=u(x)u_0(y)u_0(z)$ and 
$v(x,y,z)=v(x)v_0(y)v_0(z)$, with $u_0(y)=v_0(y)$, $u_0(z)=v_0(z)$ denoting the normalized Gaussian wave 
functions along the transverse directions. 
Similarly to the 1D initial conditions of the preceding section, the wave functions of the two 
components along the $x$-direction are given by a Thomas-Fermi approximation in the $u$ component, complementary
to the $v$ component for which we assume a Gaussian initial condition:
$v(x)= A \exp(-x^2/2w^2)$.
The dynamical evolution of the density of the 
$v$ component when considering two distinct 
initializations corresponding to a narrow 
($w=0.7~{\rm \mu m}$) and
a broad ($w=3.5~{\rm \mu m}$) Gaussian initial 
profile with amplitude $A=0.5~{\rm \mu m^{-1/2}}$ 
is demonstrated in Fig.~\ref{fig:3d}(a)--\ref{fig:3d}(h). 
Note that since the spatiotemporal evolution of the 
$u$ component is complementary to the one shown 
it is not illustrated here but it can be inferred, for 
instance, by inspecting the insets in 
Fig.~\ref{fig:3d}(a) where the instantaneous density 
profiles of both components in the $x$-$y$ plane are 
shown. 
Remarkably, the spontaneous nucleation of a
Peregrine structure along with its 
revivals takes place in this 3D, quasi-1D setting 
[Fig.~\ref{fig:3d}(a)] with its integrated, along the 
transverse $yz$ directions, density profiles
presented in Fig.~\ref{fig:3d}(c)--\ref{fig:3d}(e). 
Additionally, and as suggested by the relevant 1D 
situation, for wider initial wave packets  
a configuration reminiscent of the Christmas tree is 
generated [Fig.~\ref{fig:3d}(b)].
In both cases, very good agreement is found when fitting the analytical 1D solution to the numerically 
obtained rogue-wave pattern appearing initially at 
$t_0\approx 13.5\,$ms, but recurring at $t_1\approx 46\,$ms and $t_2\approx 81\,$ms for narrow pulses [Fig.~\ref{fig:3d}(c)-(e)] 
and around $t_3=41\,$ms, $t_4 \approx 47\,$ms and 
$t_5 \approx 53.6\,$ms for wider ones [Fig.~\ref{fig:3d}(f)-(h)]. 
The observation and persistence of the Peregrine 
structure in such a 3D setting with a quasi-1D geometry constitutes, to 
the best of our knowledge, an unprecedented result that sets the stage for an experimental 
realization of this rogue wave.

\section{Conclusions and Future Perspectives} \label{sec:conclusions}

In this work, we have presented an experimentally realizable setup to explore the formation of Peregrine solitons in repulsive two-component BECs.
First, as a proof-of-principle, we showed the formation of the Peregine soliton by direct initialization on the wave function of the minority component, and demonstrated how an effective single-component picture accurately captures the dynamics.
Additionally, we argued that in the presence of a wide external harmonic trapping potential 
Peregrine solitons can also be realized in a two-component setup, with periodic revivals stemming as a result of the presence of the trap.

We then extended our work to a more experimentally relevant situation by utilizing wide and thus effectively semiclassical Gaussian wave packets as initial conditions, both in the absence and in the presence of a parabolic trap.
In particular, by employing differently sized Gaussian profiles we were able to showcase that narrower wave packets lead to periodic revivals of a localized Peregrine-like structure resulting in a \textit{chain}like pattern.
Contrary to this dynamical evolution, broader Gaussian profiles entail the formation of a cascade of Peregrine waves, also known as a Christmas-tree structure.
Moreover, we demonstrated that in the presence of the trap the dynamics of a narrower Gaussian initial condition remains qualitatively identical to that observed in the homogeneous cases under consideration.
On the other hand, for wider Gaussian profiles we encountered a particularly interesting phenomenon that was absent in the aforementioned settings. 
While at short timescales, the dynamical evolution of both the two-component and the effective single-component models is similar to their corresponding untrapped settings, this is not the case for longer evolution times. At these longer times, dark-bright soliton-type structures are formed in the two-component setup and, contrary to the standard dark-bright solitary waves, feature unexpected \textit{very} long-time convex (up to the vicinity
of the turning point) trajectories within the parabolic trap.
Strikingly, these unprecedented ---to our knowledge--- patterns are seen to interfere anew suggesting that the system can access very long time scale recurrences of the Peregrine-like structures.

Furthermore, we attempted yet another generalization of our findings by considering also the case of mass-imbalanced mixtures.
Here, we were able to exemplify that the overall dynamical response persists for mixtures in which the minority component is the heaviest one, with the rogue and Christmas-tree pattern formation being accelerated when compared to the mass-balanced model. 
Moreover, in such mass-imbalanced mixtures we observed the emergence of dark-bright soliton-like entities that, besides oscillating within the parabolic trap, were also seen to exhibit a characteristic breathing.

Additionally, we examined the robustness of Peregrine soliton nucleation in the presence of dissipation. 
Specifically, it is demonstrated that the overall phenomenology persists for values of the dissipative parameter $\gamma$
of the order of $\sim 10^{-3}$ (i.e., temperatures of about $\sim 100\,$nK) before any signature of the Peregrine wave is lost. 
A key finding toward the realizability of our proposal in
experimental settings consists of the successful numerical realization of both the Peregrine soliton with its revivals and the Christmas-tree pattern in a 3D geometry involving a {quasi}-1D cigar-shaped harmonic trap.

Until now, the realization of a Peregrine soliton has been a challenge in the field of ultracold atoms.  
This mainly stems from difficulties to highly control experiments with attractive BECs, even more so in the presence of the modulationally unstable background that supports the Peregrine soliton.
Yet, in this work we have argued that the presence of a majority component contributes 
to the formation of more robust structures, rendering two-component {\it repulsive} BECs an appealing and potentially more suitable platform for the realization and further study of Peregrine waves in the effective form proposed herein.
At the same time, we observed the peculiar formation of dark-bright soliton-like structures 
bearing unusual oscillatory trajectories. 
The latter, along with the quantification of the oscillation period of such entities, constitute fruitful directions of study worthwhile to pursue in 
the future.
Another interesting pathway is the inclusion of three-body loss rates in order to inspect the robustness of the Peregrine wave in the long time evolution where their effect might become appreciable. 
Finally, unraveling the correlation properties of these structures when embedded in a many-body environment, e.g., as has been demonstrated for dark-bright solitons~\cite{Katsimiga2017,Katsimiga2018,Mistakidis2018}, is an intriguing topic especially so when beyond mean-field effects may come into play, potentially affecting the validity of the single-component effective description. 

\section*{Acknowledgements}

A.R.R. and P.S. acknowledge financial support by the Deutsche Forschungsgemeinschaft (DFG, German Research Foundation) -- SFB-925 -- Project No. 170620586.
G.C.K. gratefuly acknowledges financial support by the Cluster of Excellence ``Advanced Imaging of Matter'' of the Deutsche Forschungsgemeinschaft(DFG) -- EXC 2056 -- Project ID 390715994. 
S.I.M. acknowledges support from the NSF through a grant for ITAMP at Harvard University. 
This material is based upon work supported by the US National Science Foundation under Grants No. PHY-2110030 and No. DMS-1809074 (P.G.K.), as well as No. DMS-2009487 (G.B.) and No.
DMS-2106488 (B.P.).
P.G.K. is grateful to Markus Oberthaler and the Synthetic Quantum Systems group
for numerous useful discussions and insights.

\appendix 

\section{Modulational instability and structure formation in the long-time dynamics} \label{app:wedge-structure} 

In nonlinear focusing media, a small localized perturbation on a constant background can lead to a modulationally unstable region. The latter acquires a wedge-like shape characterized by a universal envelope, known as the nonlinear stage of the  modulational instability 
(see Refs.~\cite{El1993,Biondini2016b} and references therein).
For the integrable case of the NLS equation, such a region is defined by the boundaries $x_\pm=\pm4\sqrt{2P_o} t$, with $P_o$ denoting the background density.
On the other hand, beyond the integrable limit, the boundaries depend also on the nonlinearity as follows~\cite{Zakharov2013,Biondini2018a,Kraych2019}:
\begin{equation}
    x_\pm=\pm2\sqrt{-2gP_o}t.
    \label{eq:speed_limit}
\end{equation}
This expression refers to a focusing system, and thus $g<0$.
In our case, $g$ is given by Eq.~\eqref{eq:eff_g}.
\begin{figure}[t!]
    \centering
    \includegraphics[width=\linewidth]{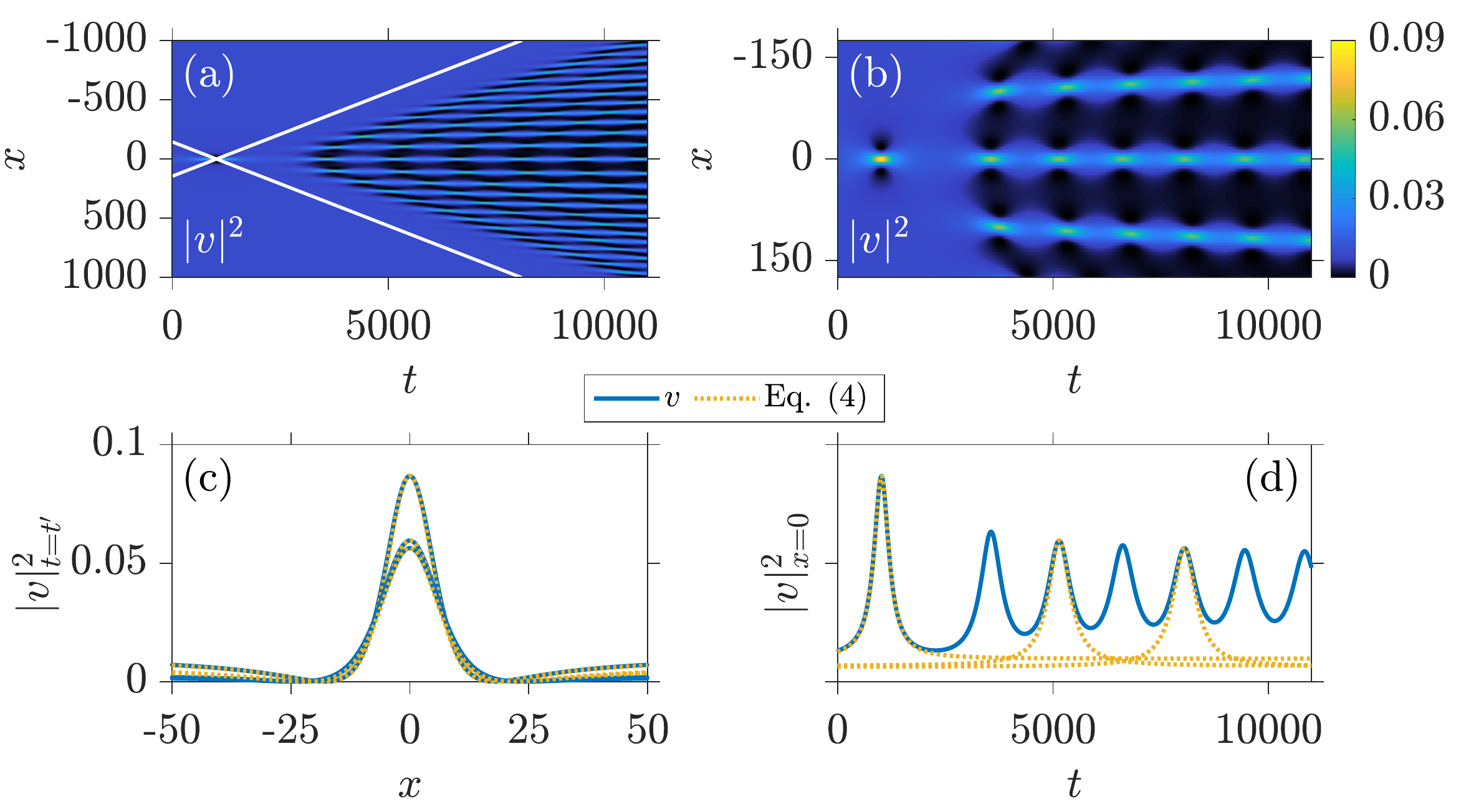}
    \caption{(a) Density evolution of the Peregrine initial condition of Eq.~\eqref{eq:peregrine}, for $P_o=0.01$ and $t_o=1000$, 
    in the homogeneous two-component setup. 
    The majority $u$ component is complementary to the minority $v$ component, and thus omitted.
    White lines, corresponding to Eq.~\eqref{eq:speed_limit}, designate the borders of the wedge-like pattern formed.
    (b) Enlargement of panel (a).
    (c) Density profile of the minority $v$ component at the time instant of Peregrine soliton formation, namely at $t_0'=1023$, $t_2'=5148$, 
    and $t_4'=8052$.
    (d) Temporal evolution of the modulation wave at $x=0$.
    In both panels (c) and (d), the corresponding analytical solutions of Eq.~\eqref{eq:peregrine} are provided for the chosen Peregrine solitons. Length and time units are expressed in terms of $a_\perp$ and $\omega_\perp^{-1}$, respectively.}
    \label{fig:wedge}
\end{figure}
In Fig.~\ref{fig:wedge}, we show the long time dynamics of a system initialized with Eq.~\eqref{eq:peregrine}, with $P_o=0.01$ and $t_o=1000$, in a repulsive two-component system (see Sec.~\ref{sec:model} for details).
The short time dynamics of this initial state is presented in Fig.~\ref{fig:peregrine_homo}.
Here, we find that the long time dynamics of such an initialization develops into a wedge-like structure similar to the one described in Ref.~\cite{Biondini2016b} and recently found experimentally in Ref.~\cite{Kraych2019}.
The analytical estimate of    Eq.~\eqref{eq:speed_limit} characterizing the boundaries of the wedge structure is illustrated in Fig.~\ref{fig:wedge}(a) with white solid lines. Evidently, a good agreement with the numerical prediction is found.

However, in this case we observe a particularly interesting behavior not reported in the previous works.
There, the wedge structure consisted of homogeneous fringes.
In our case, the fringes composing this entity are revivals of the Peregrine soliton, both in time and space. 
As such, they are 
reminiscent of the Kuznetsov-Ma soliton~\cite{Kuznetsov1977,Ma1979} and the Akhmediev breather~\cite{Akhmediev1986} and more generally 
of the doubly periodic solutions in space and time~\cite{Akhmediev1987}.

To better showcase this behavior, in Fig.~\ref{fig:wedge}(b) we present the inner region of the wedge structure.
Clearly, each of the fringes constituting the wedge structure possesses revivals of the Peregrine soliton.
However, every revival has a smaller amplitude than the previous one.
In Fig.~\ref{fig:wedge}(c), we depict several Peregrine solitons emerging at $x=0$.
In decreasing amplitude, they correspond to the initial Peregrine solitons ($t'_0=1023$), the second revival ($t'_2=5148$), and the fourth revival ($t'_4=8052$).
Additionally, we fitted Eq.~\eqref{eq:peregrine} to each of the aforementioned wave forms to verify that indeed they are Peregrine solitons.
Lastly, the temporal evolution of the density of the central fringe of the wedge 
structure is shown in Fig.~\ref{fig:wedge}(d). 
Here, each revival of the original Peregrine soliton is clearly discernible, as well as the corresponding decrease in amplitude.
Again, a fitting to the wave form of Eq.~\eqref{eq:peregrine} exhibits an adequate agreement with the selected revivals of the density, exposing their Peregrine character. 
It is important to remind the reader here that this wedge structure was predicted and found in focusing media.
In contrast, our setup corresponds to a manifestation thereof in a repulsive (defocusing) two-component BEC.

\bibliographystyle{apsrev4-1}
\bibliography{Peregrine.bib}

\begin{thebibliography}{79}%
\makeatletter
\providecommand \@ifxundefined [1]{%
 \@ifx{#1\undefined}
}%
\providecommand \@ifnum [1]{%
 \ifnum #1\expandafter \@firstoftwo
 \else \expandafter \@secondoftwo
 \fi
}%
\providecommand \@ifx [1]{%
 \ifx #1\expandafter \@firstoftwo
 \else \expandafter \@secondoftwo
 \fi
}%
\providecommand \natexlab [1]{#1}%
\providecommand \enquote  [1]{``#1''}%
\providecommand \bibnamefont  [1]{#1}%
\providecommand \bibfnamefont [1]{#1}%
\providecommand \citenamefont [1]{#1}%
\providecommand \href@noop [0]{\@secondoftwo}%
\providecommand \href [0]{\begingroup \@sanitize@url \@href}%
\providecommand \@href[1]{\@@startlink{#1}\@@href}%
\providecommand \@@href[1]{\endgroup#1\@@endlink}%
\providecommand \@sanitize@url [0]{\catcode `\\12\catcode `\$12\catcode
  `\&12\catcode `\#12\catcode `\^12\catcode `\_12\catcode `\%12\relax}%
\providecommand \@@startlink[1]{}%
\providecommand \@@endlink[0]{}%
\providecommand \url  [0]{\begingroup\@sanitize@url \@url }%
\providecommand \@url [1]{\endgroup\@href {#1}{\urlprefix }}%
\providecommand \urlprefix  [0]{URL }%
\providecommand \Eprint [0]{\href }%
\providecommand \doibase [0]{http://dx.doi.org/}%
\providecommand \selectlanguage [0]{\@gobble}%
\providecommand \bibinfo  [0]{\@secondoftwo}%
\providecommand \bibfield  [0]{\@secondoftwo}%
\providecommand \translation [1]{[#1]}%
\providecommand \BibitemOpen [0]{}%
\providecommand \bibitemStop [0]{}%
\providecommand \bibitemNoStop [0]{.\EOS\space}%
\providecommand \EOS [0]{\spacefactor3000\relax}%
\providecommand \BibitemShut  [1]{\csname bibitem#1\endcsname}%
\let\auto@bib@innerbib\@empty
\bibitem [{\citenamefont {Draper}(1966)}]{Draper1966}%
  \BibitemOpen
  \bibfield  {author} {\bibinfo {author} {\bibfnamefont {L.}~\bibnamefont
  {Draper}},\ }\href {\doibase 10.1002/j.1477-8696.1966.tb05176.x} {\bibfield
  {journal} {\bibinfo  {journal} {Weather}\ }\textbf {\bibinfo {volume} {21}},\
  \bibinfo {pages} {2} (\bibinfo {year} {1966})}\BibitemShut {NoStop}%
\bibitem [{\citenamefont {Kharif}\ \emph {et~al.}(2009)\citenamefont {Kharif},
  \citenamefont {Pelinovsky},\ and\ \citenamefont {Slunyaev}}]{Kharif2009}%
  \BibitemOpen
  \bibfield  {author} {\bibinfo {author} {\bibfnamefont {C.}~\bibnamefont
  {Kharif}}, \bibinfo {author} {\bibfnamefont {E.}~\bibnamefont {Pelinovsky}},
  \ and\ \bibinfo {author} {\bibfnamefont {A.}~\bibnamefont {Slunyaev}},\
  }\href {\doibase 10.1007/978-3-540-88419-4} {\emph {\bibinfo {title} {Rogue
  {{Waves}} in the {{Ocean}}}}},\ Advances in {{Geophysical}} and
  {{Environmental Mechanics}} and {{Mathematics}}\ (\bibinfo  {publisher}
  {{Springer-Verlag}},\ \bibinfo {address} {{Berlin Heidelberg}},\ \bibinfo
  {year} {2009})\BibitemShut {NoStop}%
\bibitem [{\citenamefont {Akhmediev}\ \emph {et~al.}(2009)\citenamefont
  {Akhmediev}, \citenamefont {{Soto-Crespo}},\ and\ \citenamefont
  {Ankiewicz}}]{Akhmediev2009}%
  \BibitemOpen
  \bibfield  {author} {\bibinfo {author} {\bibfnamefont {N.}~\bibnamefont
  {Akhmediev}}, \bibinfo {author} {\bibfnamefont {J.~M.}\ \bibnamefont
  {{Soto-Crespo}}}, \ and\ \bibinfo {author} {\bibfnamefont {A.}~\bibnamefont
  {Ankiewicz}},\ }\href {\doibase 10.1016/j.physleta.2009.04.023} {\bibfield
  {journal} {\bibinfo  {journal} {Phys. Lett. A}\ }\textbf {\bibinfo {volume}
  {373}},\ \bibinfo {pages} {2137} (\bibinfo {year} {2009})}\BibitemShut
  {NoStop}%
\bibitem [{\citenamefont {Ablowitz}\ and\ \citenamefont
  {Cole}(2021)}]{Ablowitz2021}%
  \BibitemOpen
  \bibfield  {author} {\bibinfo {author} {\bibfnamefont {M.~J.}\ \bibnamefont
  {Ablowitz}}\ and\ \bibinfo {author} {\bibfnamefont {J.~T.}\ \bibnamefont
  {Cole}},\ }\href {\doibase 10.1103/PhysRevLett.127.104101} {\bibfield
  {journal} {\bibinfo  {journal} {Phys. Rev. Lett.}\ }\textbf {\bibinfo
  {volume} {127}},\ \bibinfo {pages} {104101} (\bibinfo {year}
  {2021})}\BibitemShut {NoStop}%
\bibitem [{\citenamefont {Kuznetsov}(1977)}]{Kuznetsov1977}%
  \BibitemOpen
  \bibfield  {author} {\bibinfo {author} {\bibfnamefont {E.~A.}\ \bibnamefont
  {Kuznetsov}},\ }\href@noop {} {\bibfield  {journal} {\bibinfo  {journal}
  {Sov. Phys.-Dokl.}\ }\textbf {\bibinfo {volume} {236}},\ \bibinfo {pages}
  {575} (\bibinfo {year} {1977})}\BibitemShut {NoStop}%
\bibitem [{\citenamefont {Ma}(1979)}]{Ma1979}%
  \BibitemOpen
  \bibfield  {author} {\bibinfo {author} {\bibfnamefont {Y.-C.}\ \bibnamefont
  {Ma}},\ }\href {\doibase 10.1002/sapm197960143} {\bibfield  {journal}
  {\bibinfo  {journal} {Stud. Appl. Math.}\ }\textbf {\bibinfo {volume} {60}},\
  \bibinfo {pages} {43} (\bibinfo {year} {1979})}\BibitemShut {NoStop}%
\bibitem [{\citenamefont {Dysthe}\ and\ \citenamefont
  {Trulsen}(1999)}]{Dysthe1999}%
  \BibitemOpen
  \bibfield  {author} {\bibinfo {author} {\bibfnamefont {K.~B.}\ \bibnamefont
  {Dysthe}}\ and\ \bibinfo {author} {\bibfnamefont {K.}~\bibnamefont
  {Trulsen}},\ }\href {\doibase 10.1238/Physica.Topical.082a00048} {\bibfield
  {journal} {\bibinfo  {journal} {Phys. Scr.}\ }\textbf {\bibinfo {volume}
  {1999}},\ \bibinfo {pages} {48} (\bibinfo {year} {1999})}\BibitemShut
  {NoStop}%
\bibitem [{\citenamefont {Akhmediev}\ and\ \citenamefont
  {Korneev}(1986)}]{Akhmediev1986}%
  \BibitemOpen
  \bibfield  {author} {\bibinfo {author} {\bibfnamefont {N.~N.}\ \bibnamefont
  {Akhmediev}}\ and\ \bibinfo {author} {\bibfnamefont {V.~I.}\ \bibnamefont
  {Korneev}},\ }\href {\doibase 10.1007/BF01037866} {\bibfield  {journal}
  {\bibinfo  {journal} {Theor Math Phys}\ }\textbf {\bibinfo {volume} {69}},\
  \bibinfo {pages} {1089} (\bibinfo {year} {1986})}\BibitemShut {NoStop}%
\bibitem [{\citenamefont {Karjanto}(2020)}]{Karjanto2020}%
  \BibitemOpen
  \bibfield  {author} {\bibinfo {author} {\bibfnamefont {N.}~\bibnamefont
  {Karjanto}},\ }\href@noop {} {\bibfield  {journal} {\bibinfo  {journal}
  {ArXiv200900269 Nlin Physicsphysics}\ } (\bibinfo {year} {2020})},\ \Eprint
  {http://arxiv.org/abs/2009.00269} {arXiv:2009.00269 [nlin, physics:physics]}
  \BibitemShut {NoStop}%
\bibitem [{\citenamefont {Kibler}\ \emph {et~al.}(2010)\citenamefont {Kibler},
  \citenamefont {Fatome}, \citenamefont {Finot}, \citenamefont {Millot},
  \citenamefont {Dias}, \citenamefont {Genty}, \citenamefont {Akhmediev},\ and\
  \citenamefont {Dudley}}]{Kibler2010}%
  \BibitemOpen
  \bibfield  {author} {\bibinfo {author} {\bibfnamefont {B.}~\bibnamefont
  {Kibler}}, \bibinfo {author} {\bibfnamefont {J.}~\bibnamefont {Fatome}},
  \bibinfo {author} {\bibfnamefont {C.}~\bibnamefont {Finot}}, \bibinfo
  {author} {\bibfnamefont {G.}~\bibnamefont {Millot}}, \bibinfo {author}
  {\bibfnamefont {F.}~\bibnamefont {Dias}}, \bibinfo {author} {\bibfnamefont
  {G.}~\bibnamefont {Genty}}, \bibinfo {author} {\bibfnamefont
  {N.}~\bibnamefont {Akhmediev}}, \ and\ \bibinfo {author} {\bibfnamefont
  {J.~M.}\ \bibnamefont {Dudley}},\ }\href {\doibase 10.1038/nphys1740}
  {\bibfield  {journal} {\bibinfo  {journal} {Nature Phys}\ }\textbf {\bibinfo
  {volume} {6}},\ \bibinfo {pages} {790} (\bibinfo {year} {2010})}\BibitemShut
  {NoStop}%
\bibitem [{\citenamefont {Kibler}\ \emph {et~al.}(2012)\citenamefont {Kibler},
  \citenamefont {Fatome}, \citenamefont {Finot}, \citenamefont {Millot},
  \citenamefont {Genty}, \citenamefont {Wetzel}, \citenamefont {Akhmediev},
  \citenamefont {Dias},\ and\ \citenamefont {Dudley}}]{Kibler2012}%
  \BibitemOpen
  \bibfield  {author} {\bibinfo {author} {\bibfnamefont {B.}~\bibnamefont
  {Kibler}}, \bibinfo {author} {\bibfnamefont {J.}~\bibnamefont {Fatome}},
  \bibinfo {author} {\bibfnamefont {C.}~\bibnamefont {Finot}}, \bibinfo
  {author} {\bibfnamefont {G.}~\bibnamefont {Millot}}, \bibinfo {author}
  {\bibfnamefont {G.}~\bibnamefont {Genty}}, \bibinfo {author} {\bibfnamefont
  {B.}~\bibnamefont {Wetzel}}, \bibinfo {author} {\bibfnamefont
  {N.}~\bibnamefont {Akhmediev}}, \bibinfo {author} {\bibfnamefont
  {F.}~\bibnamefont {Dias}}, \ and\ \bibinfo {author} {\bibfnamefont {J.~M.}\
  \bibnamefont {Dudley}},\ }\href {\doibase 10.1038/srep00463} {\bibfield
  {journal} {\bibinfo  {journal} {Sci. Rep.}\ }\textbf {\bibinfo {volume}
  {2}},\ \bibinfo {pages} {463} (\bibinfo {year} {2012})}\BibitemShut {NoStop}%
\bibitem [{\citenamefont {DeVore}\ \emph {et~al.}(2013)\citenamefont {DeVore},
  \citenamefont {Solli}, \citenamefont {Borlaug}, \citenamefont {Ropers},\ and\
  \citenamefont {Jalali}}]{DeVore2013}%
  \BibitemOpen
  \bibfield  {author} {\bibinfo {author} {\bibfnamefont {P.~T.~S.}\
  \bibnamefont {DeVore}}, \bibinfo {author} {\bibfnamefont {D.~R.}\
  \bibnamefont {Solli}}, \bibinfo {author} {\bibfnamefont {D.}~\bibnamefont
  {Borlaug}}, \bibinfo {author} {\bibfnamefont {C.}~\bibnamefont {Ropers}}, \
  and\ \bibinfo {author} {\bibfnamefont {B.}~\bibnamefont {Jalali}},\ }\href
  {\doibase 10.1088/2040-8978/15/6/064001} {\bibfield  {journal} {\bibinfo
  {journal} {J. Opt.}\ }\textbf {\bibinfo {volume} {15}},\ \bibinfo {pages}
  {064001} (\bibinfo {year} {2013})}\BibitemShut {NoStop}%
\bibitem [{\citenamefont {Dudley}\ \emph {et~al.}(2014)\citenamefont {Dudley},
  \citenamefont {Dias}, \citenamefont {Erkintalo},\ and\ \citenamefont
  {Genty}}]{Dudley2014}%
  \BibitemOpen
  \bibfield  {author} {\bibinfo {author} {\bibfnamefont {J.~M.}\ \bibnamefont
  {Dudley}}, \bibinfo {author} {\bibfnamefont {F.}~\bibnamefont {Dias}},
  \bibinfo {author} {\bibfnamefont {M.}~\bibnamefont {Erkintalo}}, \ and\
  \bibinfo {author} {\bibfnamefont {G.}~\bibnamefont {Genty}},\ }\href
  {\doibase 10.1038/nphoton.2014.220} {\bibfield  {journal} {\bibinfo
  {journal} {Nature Photon}\ }\textbf {\bibinfo {volume} {8}},\ \bibinfo
  {pages} {755} (\bibinfo {year} {2014})}\BibitemShut {NoStop}%
\bibitem [{\citenamefont {Frisquet}\ \emph {et~al.}(2016)\citenamefont
  {Frisquet}, \citenamefont {Kibler}, \citenamefont {Morin}, \citenamefont
  {Baronio}, \citenamefont {Conforti}, \citenamefont {Millot},\ and\
  \citenamefont {Wabnitz}}]{Frisquet2016}%
  \BibitemOpen
  \bibfield  {author} {\bibinfo {author} {\bibfnamefont {B.}~\bibnamefont
  {Frisquet}}, \bibinfo {author} {\bibfnamefont {B.}~\bibnamefont {Kibler}},
  \bibinfo {author} {\bibfnamefont {P.}~\bibnamefont {Morin}}, \bibinfo
  {author} {\bibfnamefont {F.}~\bibnamefont {Baronio}}, \bibinfo {author}
  {\bibfnamefont {M.}~\bibnamefont {Conforti}}, \bibinfo {author}
  {\bibfnamefont {G.}~\bibnamefont {Millot}}, \ and\ \bibinfo {author}
  {\bibfnamefont {S.}~\bibnamefont {Wabnitz}},\ }\href {\doibase
  10.1038/srep20785} {\bibfield  {journal} {\bibinfo  {journal} {Sci Rep}\
  }\textbf {\bibinfo {volume} {6}},\ \bibinfo {pages} {20785} (\bibinfo {year}
  {2016})}\BibitemShut {NoStop}%
\bibitem [{\citenamefont {Sabry}\ \emph {et~al.}(2012)\citenamefont {Sabry},
  \citenamefont {Moslem},\ and\ \citenamefont {Shukla}}]{Sabry2012}%
  \BibitemOpen
  \bibfield  {author} {\bibinfo {author} {\bibfnamefont {R.}~\bibnamefont
  {Sabry}}, \bibinfo {author} {\bibfnamefont {W.~M.}\ \bibnamefont {Moslem}}, \
  and\ \bibinfo {author} {\bibfnamefont {P.~K.}\ \bibnamefont {Shukla}},\
  }\href {\doibase 10.1063/1.4772058} {\bibfield  {journal} {\bibinfo
  {journal} {Phys. Plasmas}\ }\textbf {\bibinfo {volume} {19}},\ \bibinfo
  {pages} {122903} (\bibinfo {year} {2012})}\BibitemShut {NoStop}%
\bibitem [{\citenamefont {Bains}\ \emph {et~al.}(2014)\citenamefont {Bains},
  \citenamefont {Li},\ and\ \citenamefont {Xia}}]{Bains2014}%
  \BibitemOpen
  \bibfield  {author} {\bibinfo {author} {\bibfnamefont {A.~S.}\ \bibnamefont
  {Bains}}, \bibinfo {author} {\bibfnamefont {B.}~\bibnamefont {Li}}, \ and\
  \bibinfo {author} {\bibfnamefont {L.-D.}\ \bibnamefont {Xia}},\ }\href
  {\doibase 10.1063/1.4869464} {\bibfield  {journal} {\bibinfo  {journal}
  {Phys. Plasmas}\ }\textbf {\bibinfo {volume} {21}},\ \bibinfo {pages}
  {032123} (\bibinfo {year} {2014})}\BibitemShut {NoStop}%
\bibitem [{\citenamefont {Tolba}\ \emph {et~al.}(2015)\citenamefont {Tolba},
  \citenamefont {Moslem}, \citenamefont {{El-Bedwehy}},\ and\ \citenamefont
  {{El-Labany}}}]{Tolba2015}%
  \BibitemOpen
  \bibfield  {author} {\bibinfo {author} {\bibfnamefont {R.~E.}\ \bibnamefont
  {Tolba}}, \bibinfo {author} {\bibfnamefont {W.~M.}\ \bibnamefont {Moslem}},
  \bibinfo {author} {\bibfnamefont {N.~A.}\ \bibnamefont {{El-Bedwehy}}}, \
  and\ \bibinfo {author} {\bibfnamefont {S.~K.}\ \bibnamefont {{El-Labany}}},\
  }\href {\doibase 10.1063/1.4918706} {\bibfield  {journal} {\bibinfo
  {journal} {Phys. Plasmas}\ }\textbf {\bibinfo {volume} {22}},\ \bibinfo
  {pages} {043707} (\bibinfo {year} {2015})}\BibitemShut {NoStop}%
\bibitem [{\citenamefont {Ganshin}\ \emph {et~al.}(2008)\citenamefont
  {Ganshin}, \citenamefont {Efimov}, \citenamefont {Kolmakov}, \citenamefont
  {{Mezhov-Deglin}},\ and\ \citenamefont {McClintock}}]{Ganshin2008}%
  \BibitemOpen
  \bibfield  {author} {\bibinfo {author} {\bibfnamefont {A.~N.}\ \bibnamefont
  {Ganshin}}, \bibinfo {author} {\bibfnamefont {V.~B.}\ \bibnamefont {Efimov}},
  \bibinfo {author} {\bibfnamefont {G.~V.}\ \bibnamefont {Kolmakov}}, \bibinfo
  {author} {\bibfnamefont {L.~P.}\ \bibnamefont {{Mezhov-Deglin}}}, \ and\
  \bibinfo {author} {\bibfnamefont {P.~V.~E.}\ \bibnamefont {McClintock}},\
  }\href {\doibase 10.1103/PhysRevLett.101.065303} {\bibfield  {journal}
  {\bibinfo  {journal} {Phys. Rev. Lett.}\ }\textbf {\bibinfo {volume} {101}},\
  \bibinfo {pages} {065303} (\bibinfo {year} {2008})}\BibitemShut {NoStop}%
\bibitem [{\citenamefont {Charalampidis}\ \emph
  {et~al.}(2018{\natexlab{a}})\citenamefont {Charalampidis}, \citenamefont
  {{Cuevas-Maraver}}, \citenamefont {Frantzeskakis},\ and\ \citenamefont
  {Kevrekidis}}]{Charalampidis2018}%
  \BibitemOpen
  \bibfield  {author} {\bibinfo {author} {\bibfnamefont {E.~G.}\ \bibnamefont
  {Charalampidis}}, \bibinfo {author} {\bibfnamefont {J.}~\bibnamefont
  {{Cuevas-Maraver}}}, \bibinfo {author} {\bibfnamefont {D.~J.}\ \bibnamefont
  {Frantzeskakis}}, \ and\ \bibinfo {author} {\bibfnamefont {P.~G.}\
  \bibnamefont {Kevrekidis}},\ }\href@noop {} {\bibfield  {journal} {\bibinfo
  {journal} {Rom. Rep. Phys.}\ }\textbf {\bibinfo {volume} {70}},\ \bibinfo
  {pages} {504} (\bibinfo {year} {2018}{\natexlab{a}})}\BibitemShut {NoStop}%
\bibitem [{\citenamefont {{Bakkali-Hassani}}\ \emph {et~al.}(2021)\citenamefont
  {{Bakkali-Hassani}}, \citenamefont {Maury}, \citenamefont {Zou},
  \citenamefont {Le~Cerf}, \citenamefont {{Saint-Jalm}}, \citenamefont
  {Castilho}, \citenamefont {Nascimbene}, \citenamefont {Dalibard},\ and\
  \citenamefont {Beugnon}}]{Bakkali-Hassani2021}%
  \BibitemOpen
  \bibfield  {author} {\bibinfo {author} {\bibfnamefont {B.}~\bibnamefont
  {{Bakkali-Hassani}}}, \bibinfo {author} {\bibfnamefont {C.}~\bibnamefont
  {Maury}}, \bibinfo {author} {\bibfnamefont {Y.-Q.}\ \bibnamefont {Zou}},
  \bibinfo {author} {\bibfnamefont {{\'E}.}~\bibnamefont {Le~Cerf}}, \bibinfo
  {author} {\bibfnamefont {R.}~\bibnamefont {{Saint-Jalm}}}, \bibinfo {author}
  {\bibfnamefont {P.~C.~M.}\ \bibnamefont {Castilho}}, \bibinfo {author}
  {\bibfnamefont {S.}~\bibnamefont {Nascimbene}}, \bibinfo {author}
  {\bibfnamefont {J.}~\bibnamefont {Dalibard}}, \ and\ \bibinfo {author}
  {\bibfnamefont {J.}~\bibnamefont {Beugnon}},\ }\href {\doibase
  10.1103/PhysRevLett.127.023603} {\bibfield  {journal} {\bibinfo  {journal}
  {Phys. Rev. Lett.}\ }\textbf {\bibinfo {volume} {127}},\ \bibinfo {pages}
  {023603} (\bibinfo {year} {2021})}\BibitemShut {NoStop}%
\bibitem [{\citenamefont {Yan}(2012)}]{Yan2012a}%
  \BibitemOpen
  \bibfield  {author} {\bibinfo {author} {\bibfnamefont {Z.}~\bibnamefont
  {Yan}},\ }\href {\doibase 10.1088/1742-6596/400/1/012084} {\bibfield
  {journal} {\bibinfo  {journal} {J. Phys.: Conf. Ser.}\ }\textbf {\bibinfo
  {volume} {400}},\ \bibinfo {pages} {012084} (\bibinfo {year}
  {2012})}\BibitemShut {NoStop}%
\bibitem [{\citenamefont {Onorato}\ \emph {et~al.}(2013)\citenamefont
  {Onorato}, \citenamefont {Residori}, \citenamefont {Bortolozzo},
  \citenamefont {Montina},\ and\ \citenamefont {Arecchi}}]{Onorato2013}%
  \BibitemOpen
  \bibfield  {author} {\bibinfo {author} {\bibfnamefont {M.}~\bibnamefont
  {Onorato}}, \bibinfo {author} {\bibfnamefont {S.}~\bibnamefont {Residori}},
  \bibinfo {author} {\bibfnamefont {U.}~\bibnamefont {Bortolozzo}}, \bibinfo
  {author} {\bibfnamefont {A.}~\bibnamefont {Montina}}, \ and\ \bibinfo
  {author} {\bibfnamefont {F.~T.}\ \bibnamefont {Arecchi}},\ }\href {\doibase
  10.1016/j.physrep.2013.03.001} {\bibfield  {journal} {\bibinfo  {journal}
  {Phys. Rep.}\ }\textbf {\bibinfo {volume} {528}},\ \bibinfo {pages} {47}
  (\bibinfo {year} {2013})}\BibitemShut {NoStop}%
\bibitem [{\citenamefont {Mihalache}(2017)}]{Mihalache2017}%
  \BibitemOpen
  \bibfield  {author} {\bibinfo {author} {\bibfnamefont {D.}~\bibnamefont
  {Mihalache}},\ }\href@noop {} {\bibfield  {journal} {\bibinfo  {journal}
  {Rom. Rep. Phys}\ }\textbf {\bibinfo {volume} {69}},\ \bibinfo {pages} {28}
  (\bibinfo {year} {2017})}\BibitemShut {NoStop}%
\bibitem [{\citenamefont {Peregrine}(1983)}]{Peregrine1983}%
  \BibitemOpen
  \bibfield  {author} {\bibinfo {author} {\bibfnamefont {D.~H.}\ \bibnamefont
  {Peregrine}},\ }\href {\doibase 10.1017/S0334270000003891} {\bibfield
  {journal} {\bibinfo  {journal} {ANZIAM J.}\ }\textbf {\bibinfo {volume}
  {25}},\ \bibinfo {pages} {16} (\bibinfo {year} {1983})}\BibitemShut {NoStop}%
\bibitem [{\citenamefont {Chabchoub}\ \emph {et~al.}(2011)\citenamefont
  {Chabchoub}, \citenamefont {Hoffmann},\ and\ \citenamefont
  {Akhmediev}}]{Chabchoub2011}%
  \BibitemOpen
  \bibfield  {author} {\bibinfo {author} {\bibfnamefont {A.}~\bibnamefont
  {Chabchoub}}, \bibinfo {author} {\bibfnamefont {N.~P.}\ \bibnamefont
  {Hoffmann}}, \ and\ \bibinfo {author} {\bibfnamefont {N.}~\bibnamefont
  {Akhmediev}},\ }\href {\doibase 10.1103/PhysRevLett.106.204502} {\bibfield
  {journal} {\bibinfo  {journal} {Phys. Rev. Lett.}\ }\textbf {\bibinfo
  {volume} {106}},\ \bibinfo {pages} {204502} (\bibinfo {year}
  {2011})}\BibitemShut {NoStop}%
\bibitem [{\citenamefont {Chabchoub}\ \emph {et~al.}(2012)\citenamefont
  {Chabchoub}, \citenamefont {Hoffmann}, \citenamefont {Onorato},\ and\
  \citenamefont {Akhmediev}}]{Chabchoub2012}%
  \BibitemOpen
  \bibfield  {author} {\bibinfo {author} {\bibfnamefont {A.}~\bibnamefont
  {Chabchoub}}, \bibinfo {author} {\bibfnamefont {N.}~\bibnamefont {Hoffmann}},
  \bibinfo {author} {\bibfnamefont {M.}~\bibnamefont {Onorato}}, \ and\
  \bibinfo {author} {\bibfnamefont {N.}~\bibnamefont {Akhmediev}},\ }\href
  {\doibase 10.1103/PhysRevX.2.011015} {\bibfield  {journal} {\bibinfo
  {journal} {Phys. Rev. X}\ }\textbf {\bibinfo {volume} {2}},\ \bibinfo {pages}
  {011015} (\bibinfo {year} {2012})}\BibitemShut {NoStop}%
\bibitem [{\citenamefont {Chabchoub}\ and\ \citenamefont
  {Fink}(2014)}]{Chabchoub2014}%
  \BibitemOpen
  \bibfield  {author} {\bibinfo {author} {\bibfnamefont {A.}~\bibnamefont
  {Chabchoub}}\ and\ \bibinfo {author} {\bibfnamefont {M.}~\bibnamefont
  {Fink}},\ }\href {\doibase 10.1103/PhysRevLett.112.124101} {\bibfield
  {journal} {\bibinfo  {journal} {Phys. Rev. Lett.}\ }\textbf {\bibinfo
  {volume} {112}},\ \bibinfo {pages} {124101} (\bibinfo {year}
  {2014})}\BibitemShut {NoStop}%
\bibitem [{\citenamefont {Tikan}\ \emph {et~al.}(2021)\citenamefont {Tikan},
  \citenamefont {Bonnefoy}, \citenamefont {Roberti}, \citenamefont {El},
  \citenamefont {Tovbis}, \citenamefont {Ducrozet}, \citenamefont {Cazaubiel},
  \citenamefont {Prabhudesai}, \citenamefont {Michel}, \citenamefont {Copie},
  \citenamefont {Falcon}, \citenamefont {Randoux},\ and\ \citenamefont
  {Suret}}]{Tikan2021}%
  \BibitemOpen
  \bibfield  {author} {\bibinfo {author} {\bibfnamefont {A.}~\bibnamefont
  {Tikan}}, \bibinfo {author} {\bibfnamefont {F.}~\bibnamefont {Bonnefoy}},
  \bibinfo {author} {\bibfnamefont {G.}~\bibnamefont {Roberti}}, \bibinfo
  {author} {\bibfnamefont {G.}~\bibnamefont {El}}, \bibinfo {author}
  {\bibfnamefont {A.}~\bibnamefont {Tovbis}}, \bibinfo {author} {\bibfnamefont
  {G.}~\bibnamefont {Ducrozet}}, \bibinfo {author} {\bibfnamefont
  {A.}~\bibnamefont {Cazaubiel}}, \bibinfo {author} {\bibfnamefont
  {G.}~\bibnamefont {Prabhudesai}}, \bibinfo {author} {\bibfnamefont
  {G.}~\bibnamefont {Michel}}, \bibinfo {author} {\bibfnamefont
  {F.}~\bibnamefont {Copie}}, \bibinfo {author} {\bibfnamefont
  {E.}~\bibnamefont {Falcon}}, \bibinfo {author} {\bibfnamefont
  {S.}~\bibnamefont {Randoux}}, \ and\ \bibinfo {author} {\bibfnamefont
  {P.}~\bibnamefont {Suret}},\ }\href@noop {} {\bibfield  {journal} {\bibinfo
  {journal} {ArXiv210802698 Nlin Physicsphysics}\ } (\bibinfo {year} {2021})},\
  \Eprint {http://arxiv.org/abs/2108.02698} {arXiv:2108.02698 [nlin,
  physics:physics]} \BibitemShut {NoStop}%
\bibitem [{\citenamefont {Bailung}\ \emph {et~al.}(2011)\citenamefont
  {Bailung}, \citenamefont {Sharma},\ and\ \citenamefont
  {Nakamura}}]{Bailung2011}%
  \BibitemOpen
  \bibfield  {author} {\bibinfo {author} {\bibfnamefont {H.}~\bibnamefont
  {Bailung}}, \bibinfo {author} {\bibfnamefont {S.~K.}\ \bibnamefont {Sharma}},
  \ and\ \bibinfo {author} {\bibfnamefont {Y.}~\bibnamefont {Nakamura}},\
  }\href {\doibase 10.1103/PhysRevLett.107.255005} {\bibfield  {journal}
  {\bibinfo  {journal} {Phys. Rev. Lett.}\ }\textbf {\bibinfo {volume} {107}},\
  \bibinfo {pages} {255005} (\bibinfo {year} {2011})}\BibitemShut {NoStop}%
\bibitem [{\citenamefont {Tikan}\ \emph {et~al.}(2017)\citenamefont {Tikan},
  \citenamefont {Billet}, \citenamefont {El}, \citenamefont {Tovbis},
  \citenamefont {Bertola}, \citenamefont {Sylvestre}, \citenamefont {Gustave},
  \citenamefont {Randoux}, \citenamefont {Genty}, \citenamefont {Suret},\ and\
  \citenamefont {Dudley}}]{Tikan2017}%
  \BibitemOpen
  \bibfield  {author} {\bibinfo {author} {\bibfnamefont {A.}~\bibnamefont
  {Tikan}}, \bibinfo {author} {\bibfnamefont {C.}~\bibnamefont {Billet}},
  \bibinfo {author} {\bibfnamefont {G.}~\bibnamefont {El}}, \bibinfo {author}
  {\bibfnamefont {A.}~\bibnamefont {Tovbis}}, \bibinfo {author} {\bibfnamefont
  {M.}~\bibnamefont {Bertola}}, \bibinfo {author} {\bibfnamefont
  {T.}~\bibnamefont {Sylvestre}}, \bibinfo {author} {\bibfnamefont
  {F.}~\bibnamefont {Gustave}}, \bibinfo {author} {\bibfnamefont
  {S.}~\bibnamefont {Randoux}}, \bibinfo {author} {\bibfnamefont
  {G.}~\bibnamefont {Genty}}, \bibinfo {author} {\bibfnamefont
  {P.}~\bibnamefont {Suret}}, \ and\ \bibinfo {author} {\bibfnamefont {J.~M.}\
  \bibnamefont {Dudley}},\ }\href {\doibase 10.1103/PhysRevLett.119.033901}
  {\bibfield  {journal} {\bibinfo  {journal} {Phys. Rev. Lett.}\ }\textbf
  {\bibinfo {volume} {119}},\ \bibinfo {pages} {033901} (\bibinfo {year}
  {2017})}\BibitemShut {NoStop}%
\bibitem [{\citenamefont {Marcucci}\ \emph {et~al.}(2019)\citenamefont
  {Marcucci}, \citenamefont {Pierangeli}, \citenamefont {Agranat},
  \citenamefont {Lee}, \citenamefont {DelRe},\ and\ \citenamefont
  {Conti}}]{Marcucci2019}%
  \BibitemOpen
  \bibfield  {author} {\bibinfo {author} {\bibfnamefont {G.}~\bibnamefont
  {Marcucci}}, \bibinfo {author} {\bibfnamefont {D.}~\bibnamefont
  {Pierangeli}}, \bibinfo {author} {\bibfnamefont {A.~J.}\ \bibnamefont
  {Agranat}}, \bibinfo {author} {\bibfnamefont {R.-K.}\ \bibnamefont {Lee}},
  \bibinfo {author} {\bibfnamefont {E.}~\bibnamefont {DelRe}}, \ and\ \bibinfo
  {author} {\bibfnamefont {C.}~\bibnamefont {Conti}},\ }\href {\doibase
  10.1038/s41467-019-12815-0} {\bibfield  {journal} {\bibinfo  {journal} {Nat.
  Commun.}\ }\textbf {\bibinfo {volume} {10}},\ \bibinfo {pages} {5090}
  (\bibinfo {year} {2019})}\BibitemShut {NoStop}%
\bibitem [{\citenamefont {Chowdury}\ and\ \citenamefont
  {Chang}(2021)}]{Chowdury2021}%
  \BibitemOpen
  \bibfield  {author} {\bibinfo {author} {\bibfnamefont {A.}~\bibnamefont
  {Chowdury}}\ and\ \bibinfo {author} {\bibfnamefont {W.}~\bibnamefont
  {Chang}},\ }\href {\doibase 10.1103/PhysRevResearch.3.L032060} {\bibfield
  {journal} {\bibinfo  {journal} {Phys. Rev. Research}\ }\textbf {\bibinfo
  {volume} {3}},\ \bibinfo {pages} {L032060} (\bibinfo {year}
  {2021})}\BibitemShut {NoStop}%
\bibitem [{\citenamefont {Pitaevskii}\ and\ \citenamefont
  {Stringari}(2003)}]{Pitaevskii2003}%
  \BibitemOpen
  \bibfield  {author} {\bibinfo {author} {\bibfnamefont {L.~P.}\ \bibnamefont
  {Pitaevskii}}\ and\ \bibinfo {author} {\bibfnamefont {S.}~\bibnamefont
  {Stringari}},\ }\href@noop {} {\emph {\bibinfo {title} {Bose-{{Einstein}}
  Condensation}}},\ \bibinfo {series} {Oxford Science Publications}\ No.\
  \bibinfo {number} {116}\ (\bibinfo  {publisher} {{Clarendon Press}},\
  \bibinfo {address} {{Oxford ; New York}},\ \bibinfo {year}
  {2003})\BibitemShut {NoStop}%
\bibitem [{\citenamefont {Zhao}(2013)}]{Zhao2013a}%
  \BibitemOpen
  \bibfield  {author} {\bibinfo {author} {\bibfnamefont {L.-C.}\ \bibnamefont
  {Zhao}},\ }\href {\doibase 10.1016/j.aop.2012.10.010} {\bibfield  {journal}
  {\bibinfo  {journal} {Annals of Physics}\ }\textbf {\bibinfo {volume}
  {329}},\ \bibinfo {pages} {73} (\bibinfo {year} {2013})}\BibitemShut
  {NoStop}%
\bibitem [{\citenamefont {Li}\ \emph {et~al.}(2018)\citenamefont {Li},
  \citenamefont {Huo}, \citenamefont {Li}, \citenamefont {He},\ and\
  \citenamefont {Xu}}]{Li2018}%
  \BibitemOpen
  \bibfield  {author} {\bibinfo {author} {\bibfnamefont {Z.-D.}\ \bibnamefont
  {Li}}, \bibinfo {author} {\bibfnamefont {C.-Z.}\ \bibnamefont {Huo}},
  \bibinfo {author} {\bibfnamefont {Q.-Y.}\ \bibnamefont {Li}}, \bibinfo
  {author} {\bibfnamefont {P.-B.}\ \bibnamefont {He}}, \ and\ \bibinfo {author}
  {\bibfnamefont {T.-F.}\ \bibnamefont {Xu}},\ }\href {\doibase
  10.1088/1674-1056/27/4/040505} {\bibfield  {journal} {\bibinfo  {journal}
  {Chinese Phys. B}\ }\textbf {\bibinfo {volume} {27}},\ \bibinfo {pages}
  {040505} (\bibinfo {year} {2018})}\BibitemShut {NoStop}%
\bibitem [{\citenamefont {Chaachoua~Sameut}\ \emph {et~al.}(2020)\citenamefont
  {Chaachoua~Sameut}, \citenamefont {Pattu}, \citenamefont {Al~Khawaja},
  \citenamefont {Benarous},\ and\ \citenamefont
  {Belkroukra}}]{ChaachouaSameut2020}%
  \BibitemOpen
  \bibfield  {author} {\bibinfo {author} {\bibfnamefont {H.}~\bibnamefont
  {Chaachoua~Sameut}}, \bibinfo {author} {\bibfnamefont {S.}~\bibnamefont
  {Pattu}}, \bibinfo {author} {\bibfnamefont {U.}~\bibnamefont {Al~Khawaja}},
  \bibinfo {author} {\bibfnamefont {M.}~\bibnamefont {Benarous}}, \ and\
  \bibinfo {author} {\bibfnamefont {H.}~\bibnamefont {Belkroukra}},\ }\href
  {\doibase 10.3103/S1541308X20030036} {\bibfield  {journal} {\bibinfo
  {journal} {Phys. Wave Phen.}\ }\textbf {\bibinfo {volume} {28}},\ \bibinfo
  {pages} {305} (\bibinfo {year} {2020})}\BibitemShut {NoStop}%
\bibitem [{\citenamefont {K{\"o}hler}\ \emph {et~al.}(2006)\citenamefont
  {K{\"o}hler}, \citenamefont {G{\'o}ral},\ and\ \citenamefont
  {Julienne}}]{Kohler2006}%
  \BibitemOpen
  \bibfield  {author} {\bibinfo {author} {\bibfnamefont {T.}~\bibnamefont
  {K{\"o}hler}}, \bibinfo {author} {\bibfnamefont {K.}~\bibnamefont
  {G{\'o}ral}}, \ and\ \bibinfo {author} {\bibfnamefont {P.~S.}\ \bibnamefont
  {Julienne}},\ }\href {\doibase 10.1103/RevModPhys.78.1311} {\bibfield
  {journal} {\bibinfo  {journal} {Rev. Mod. Phys.}\ }\textbf {\bibinfo {volume}
  {78}},\ \bibinfo {pages} {1311} (\bibinfo {year} {2006})}\BibitemShut
  {NoStop}%
\bibitem [{\citenamefont {Chin}\ \emph {et~al.}(2010)\citenamefont {Chin},
  \citenamefont {Grimm}, \citenamefont {Julienne},\ and\ \citenamefont
  {Tiesinga}}]{Chin2010}%
  \BibitemOpen
  \bibfield  {author} {\bibinfo {author} {\bibfnamefont {C.}~\bibnamefont
  {Chin}}, \bibinfo {author} {\bibfnamefont {R.}~\bibnamefont {Grimm}},
  \bibinfo {author} {\bibfnamefont {P.}~\bibnamefont {Julienne}}, \ and\
  \bibinfo {author} {\bibfnamefont {E.}~\bibnamefont {Tiesinga}},\ }\href
  {\doibase 10.1103/RevModPhys.82.1225} {\bibfield  {journal} {\bibinfo
  {journal} {Rev. Mod. Phys.}\ }\textbf {\bibinfo {volume} {82}},\ \bibinfo
  {pages} {1225} (\bibinfo {year} {2010})}\BibitemShut {NoStop}%
\bibitem [{\citenamefont {Olshanii}(1998)}]{Olshanii1998}%
  \BibitemOpen
  \bibfield  {author} {\bibinfo {author} {\bibfnamefont {M.}~\bibnamefont
  {Olshanii}},\ }\href {\doibase 10.1103/PhysRevLett.81.938} {\bibfield
  {journal} {\bibinfo  {journal} {Phys. Rev. Lett.}\ }\textbf {\bibinfo
  {volume} {81}},\ \bibinfo {pages} {938} (\bibinfo {year} {1998})}\BibitemShut
  {NoStop}%
\bibitem [{\citenamefont {Kim}\ \emph {et~al.}(2006)\citenamefont {Kim},
  \citenamefont {Melezhik},\ and\ \citenamefont {Schmelcher}}]{Kim2006}%
  \BibitemOpen
  \bibfield  {author} {\bibinfo {author} {\bibfnamefont {J.~I.}\ \bibnamefont
  {Kim}}, \bibinfo {author} {\bibfnamefont {V.~S.}\ \bibnamefont {Melezhik}}, \
  and\ \bibinfo {author} {\bibfnamefont {P.}~\bibnamefont {Schmelcher}},\
  }\href {\doibase 10.1103/PhysRevLett.97.193203} {\bibfield  {journal}
  {\bibinfo  {journal} {Phys. Rev. Lett.}\ }\textbf {\bibinfo {volume} {97}},\
  \bibinfo {pages} {193203} (\bibinfo {year} {2006})}\BibitemShut {NoStop}%
\bibitem [{\citenamefont {Henderson}\ \emph {et~al.}(2009)\citenamefont
  {Henderson}, \citenamefont {Ryu}, \citenamefont {MacCormick},\ and\
  \citenamefont {Boshier}}]{Henderson2009}%
  \BibitemOpen
  \bibfield  {author} {\bibinfo {author} {\bibfnamefont {K.}~\bibnamefont
  {Henderson}}, \bibinfo {author} {\bibfnamefont {C.}~\bibnamefont {Ryu}},
  \bibinfo {author} {\bibfnamefont {C.}~\bibnamefont {MacCormick}}, \ and\
  \bibinfo {author} {\bibfnamefont {M.~G.}\ \bibnamefont {Boshier}},\ }\href
  {\doibase 10.1088/1367-2630/11/4/043030} {\bibfield  {journal} {\bibinfo
  {journal} {New J. Phys.}\ }\textbf {\bibinfo {volume} {11}},\ \bibinfo
  {pages} {043030} (\bibinfo {year} {2009})}\BibitemShut {NoStop}%
\bibitem [{\citenamefont {Grimm}\ \emph {et~al.}(2000)\citenamefont {Grimm},
  \citenamefont {Weidem{\"u}ller},\ and\ \citenamefont
  {Ovchinnikov}}]{Grimm2000}%
  \BibitemOpen
  \bibfield  {author} {\bibinfo {author} {\bibfnamefont {R.}~\bibnamefont
  {Grimm}}, \bibinfo {author} {\bibfnamefont {M.}~\bibnamefont
  {Weidem{\"u}ller}}, \ and\ \bibinfo {author} {\bibfnamefont {Y.~B.}\
  \bibnamefont {Ovchinnikov}},\ }\href {\doibase 10.1016/S1049-250X(08)60186-X}
  {\bibfield  {journal} {\bibinfo  {journal} {Adv. At. Mol. Opt. Phys.}\
  }\textbf {\bibinfo {volume} {42}},\ \bibinfo {pages} {95} (\bibinfo {year}
  {2000})}\BibitemShut {NoStop}%
\bibitem [{\citenamefont {Chiao}\ \emph {et~al.}(1964)\citenamefont {Chiao},
  \citenamefont {Garmire},\ and\ \citenamefont
  {Townes}}]{Chiao1964SelfTrapping}%
  \BibitemOpen
  \bibfield  {author} {\bibinfo {author} {\bibfnamefont {R.~Y.}\ \bibnamefont
  {Chiao}}, \bibinfo {author} {\bibfnamefont {E.}~\bibnamefont {Garmire}}, \
  and\ \bibinfo {author} {\bibfnamefont {C.~H.}\ \bibnamefont {Townes}},\
  }\href {\doibase 10.1103/PhysRevLett.13.479} {\bibfield  {journal} {\bibinfo
  {journal} {Phys. Rev. Lett.}\ }\textbf {\bibinfo {volume} {13}},\ \bibinfo
  {pages} {479} (\bibinfo {year} {1964})}\BibitemShut {NoStop}%
\bibitem [{\citenamefont {Dutton}\ and\ \citenamefont
  {Clark}(2005)}]{Dutton2005}%
  \BibitemOpen
  \bibfield  {author} {\bibinfo {author} {\bibfnamefont {Z.}~\bibnamefont
  {Dutton}}\ and\ \bibinfo {author} {\bibfnamefont {C.~W.}\ \bibnamefont
  {Clark}},\ }\href {\doibase 10.1103/PhysRevA.71.063618} {\bibfield  {journal}
  {\bibinfo  {journal} {Phys. Rev. A}\ }\textbf {\bibinfo {volume} {71}},\
  \bibinfo {pages} {063618} (\bibinfo {year} {2005})}\BibitemShut {NoStop}%
\bibitem [{Note1()}]{Note1}%
  \BibitemOpen
  \bibinfo {note} {It is relevant to note as an aside here that the Townes
  soliton has also been created via a different method very recently in~\cite
  {Chen2021}; the latter method involved the modulational instability of a
  bright solitonic stripe in two-dimensional space.}\BibitemShut {Stop}%
\bibitem [{\citenamefont {Kevrekidis}\ \emph {et~al.}(2015)\citenamefont
  {Kevrekidis}, \citenamefont {Frantzeskakis},\ and\ \citenamefont
  {{Carretero-Gonz{\'a}lez}}}]{Kevrekidis2015}%
  \BibitemOpen
  \bibfield  {author} {\bibinfo {author} {\bibfnamefont {P.~G.}\ \bibnamefont
  {Kevrekidis}}, \bibinfo {author} {\bibfnamefont {D.~J.}\ \bibnamefont
  {Frantzeskakis}}, \ and\ \bibinfo {author} {\bibfnamefont {R.}~\bibnamefont
  {{Carretero-Gonz{\'a}lez}}},\ }\href {\doibase 10.1137/1.9781611973945}
  {\emph {\bibinfo {title} {The {{Defocusing Nonlinear Schr\"odinger
  Equation}}: From Dark Soliton to Vortices and Vortex Rings}}},\ Other
  {{Titles}} in {{Applied Mathematics}}\ (\bibinfo  {publisher} {{Society for
  Industrial and Applied Mathematics}},\ \bibinfo {address} {{Philadelphia}},\
  \bibinfo {year} {2015})\BibitemShut {NoStop}%
\bibitem [{\citenamefont {Bertola}\ and\ \citenamefont
  {Tovbis}(2013)}]{Bertola2013}%
  \BibitemOpen
  \bibfield  {author} {\bibinfo {author} {\bibfnamefont {M.}~\bibnamefont
  {Bertola}}\ and\ \bibinfo {author} {\bibfnamefont {A.}~\bibnamefont
  {Tovbis}},\ }\href {\doibase 10.1002/cpa.21445} {\bibfield  {journal}
  {\bibinfo  {journal} {Commun. Pure Appl. Math.}\ }\textbf {\bibinfo {volume}
  {66}},\ \bibinfo {pages} {678} (\bibinfo {year} {2013})}\BibitemShut
  {NoStop}%
\bibitem [{\citenamefont {Lannig}\ \emph {et~al.}(2020)\citenamefont {Lannig},
  \citenamefont {Schmied}, \citenamefont {Pr{\"u}fer}, \citenamefont {Kunkel},
  \citenamefont {Strohmaier}, \citenamefont {Strobel}, \citenamefont
  {Gasenzer}, \citenamefont {Kevrekidis},\ and\ \citenamefont
  {Oberthaler}}]{Lannig2020Collisions}%
  \BibitemOpen
  \bibfield  {author} {\bibinfo {author} {\bibfnamefont {S.}~\bibnamefont
  {Lannig}}, \bibinfo {author} {\bibfnamefont {C.-M.}\ \bibnamefont {Schmied}},
  \bibinfo {author} {\bibfnamefont {M.}~\bibnamefont {Pr{\"u}fer}}, \bibinfo
  {author} {\bibfnamefont {P.}~\bibnamefont {Kunkel}}, \bibinfo {author}
  {\bibfnamefont {R.}~\bibnamefont {Strohmaier}}, \bibinfo {author}
  {\bibfnamefont {H.}~\bibnamefont {Strobel}}, \bibinfo {author} {\bibfnamefont
  {T.}~\bibnamefont {Gasenzer}}, \bibinfo {author} {\bibfnamefont {P.~G.}\
  \bibnamefont {Kevrekidis}}, \ and\ \bibinfo {author} {\bibfnamefont {M.~K.}\
  \bibnamefont {Oberthaler}},\ }\href {\doibase 10.1103/PhysRevLett.125.170401}
  {\bibfield  {journal} {\bibinfo  {journal} {Phys. Rev. Lett.}\ }\textbf
  {\bibinfo {volume} {125}},\ \bibinfo {pages} {170401} (\bibinfo {year}
  {2020})}\BibitemShut {NoStop}%
\bibitem [{\citenamefont {Proukakis}\ and\ \citenamefont
  {Jackson}(2008)}]{Proukakis2008}%
  \BibitemOpen
  \bibfield  {author} {\bibinfo {author} {\bibfnamefont {N.~P.}\ \bibnamefont
  {Proukakis}}\ and\ \bibinfo {author} {\bibfnamefont {B.}~\bibnamefont
  {Jackson}},\ }\href {\doibase 10.1088/0953-4075/41/20/203002} {\bibfield
  {journal} {\bibinfo  {journal} {J. Phys. B: At. Mol. Opt. Phys.}\ }\textbf
  {\bibinfo {volume} {41}},\ \bibinfo {pages} {203002} (\bibinfo {year}
  {2008})}\BibitemShut {NoStop}%
\bibitem [{\citenamefont {Katsimiga}\ \emph {et~al.}(2021)\citenamefont
  {Katsimiga}, \citenamefont {Mistakidis}, \citenamefont {Schmelcher},\ and\
  \citenamefont {Kevrekidis}}]{Katsimiga2021Phase}%
  \BibitemOpen
  \bibfield  {author} {\bibinfo {author} {\bibfnamefont {G.~C.}\ \bibnamefont
  {Katsimiga}}, \bibinfo {author} {\bibfnamefont {S.~I.}\ \bibnamefont
  {Mistakidis}}, \bibinfo {author} {\bibfnamefont {P.}~\bibnamefont
  {Schmelcher}}, \ and\ \bibinfo {author} {\bibfnamefont {P.~G.}\ \bibnamefont
  {Kevrekidis}},\ }\href {\doibase 10.1088/1367-2630/abd27c} {\bibfield
  {journal} {\bibinfo  {journal} {New J. Phys.}\ }\textbf {\bibinfo {volume}
  {23}},\ \bibinfo {pages} {013015} (\bibinfo {year} {2021})}\BibitemShut
  {NoStop}%
\bibitem [{\citenamefont {Biondini}\ and\ \citenamefont
  {Mantzavinos}(2016)}]{Biondini2016b}%
  \BibitemOpen
  \bibfield  {author} {\bibinfo {author} {\bibfnamefont {G.}~\bibnamefont
  {Biondini}}\ and\ \bibinfo {author} {\bibfnamefont {D.}~\bibnamefont
  {Mantzavinos}},\ }\href {\doibase 10.1103/PhysRevLett.116.043902} {\bibfield
  {journal} {\bibinfo  {journal} {Phys. Rev. Lett.}\ }\textbf {\bibinfo
  {volume} {116}},\ \bibinfo {pages} {043902} (\bibinfo {year}
  {2016})}\BibitemShut {NoStop}%
\bibitem [{\citenamefont {Kraych}\ \emph {et~al.}(2019)\citenamefont {Kraych},
  \citenamefont {Suret}, \citenamefont {El},\ and\ \citenamefont
  {Randoux}}]{Kraych2019}%
  \BibitemOpen
  \bibfield  {author} {\bibinfo {author} {\bibfnamefont {A.~E.}\ \bibnamefont
  {Kraych}}, \bibinfo {author} {\bibfnamefont {P.}~\bibnamefont {Suret}},
  \bibinfo {author} {\bibfnamefont {G.}~\bibnamefont {El}}, \ and\ \bibinfo
  {author} {\bibfnamefont {S.}~\bibnamefont {Randoux}},\ }\href {\doibase
  10.1103/PhysRevLett.122.054101} {\bibfield  {journal} {\bibinfo  {journal}
  {Phys. Rev. Lett.}\ }\textbf {\bibinfo {volume} {122}},\ \bibinfo {pages}
  {054101} (\bibinfo {year} {2019})}\BibitemShut {NoStop}%
\bibitem [{\citenamefont {Pethick}\ and\ \citenamefont
  {Smith}(2008)}]{Pethick2008}%
  \BibitemOpen
  \bibfield  {author} {\bibinfo {author} {\bibfnamefont {C.~J.}\ \bibnamefont
  {Pethick}}\ and\ \bibinfo {author} {\bibfnamefont {H.}~\bibnamefont
  {Smith}},\ }\href {\doibase 10.1017/CBO9780511802850} {\emph {\bibinfo
  {title} {Bose\textendash{{Einstein Condensation}} in {{Dilute Gases}}}}},\
  \bibinfo {edition} {2nd}\ ed.\ (\bibinfo  {publisher} {{Cambridge University
  Press}},\ \bibinfo {address} {{Cambridge}},\ \bibinfo {year}
  {2008})\BibitemShut {NoStop}%
\bibitem [{\citenamefont {Egorov}\ \emph {et~al.}(2013)\citenamefont {Egorov},
  \citenamefont {Opanchuk}, \citenamefont {Drummond}, \citenamefont {Hall},
  \citenamefont {Hannaford},\ and\ \citenamefont {Sidorov}}]{Egorov2013}%
  \BibitemOpen
  \bibfield  {author} {\bibinfo {author} {\bibfnamefont {M.}~\bibnamefont
  {Egorov}}, \bibinfo {author} {\bibfnamefont {B.}~\bibnamefont {Opanchuk}},
  \bibinfo {author} {\bibfnamefont {P.}~\bibnamefont {Drummond}}, \bibinfo
  {author} {\bibfnamefont {B.~V.}\ \bibnamefont {Hall}}, \bibinfo {author}
  {\bibfnamefont {P.}~\bibnamefont {Hannaford}}, \ and\ \bibinfo {author}
  {\bibfnamefont {A.~I.}\ \bibnamefont {Sidorov}},\ }\href {\doibase
  10.1103/PhysRevA.87.053614} {\bibfield  {journal} {\bibinfo  {journal} {Phys.
  Rev. A}\ }\textbf {\bibinfo {volume} {87}},\ \bibinfo {pages} {053614}
  (\bibinfo {year} {2013})}\BibitemShut {NoStop}%
\bibitem [{\citenamefont {Johnson}\ \emph {et~al.}(2012)\citenamefont
  {Johnson}, \citenamefont {Bruderer}, \citenamefont {Cai}, \citenamefont
  {Clark}, \citenamefont {Bao},\ and\ \citenamefont
  {Jaksch}}]{Johnson2012Breathing}%
  \BibitemOpen
  \bibfield  {author} {\bibinfo {author} {\bibfnamefont {T.~H.}\ \bibnamefont
  {Johnson}}, \bibinfo {author} {\bibfnamefont {M.}~\bibnamefont {Bruderer}},
  \bibinfo {author} {\bibfnamefont {Y.}~\bibnamefont {Cai}}, \bibinfo {author}
  {\bibfnamefont {S.~R.}\ \bibnamefont {Clark}}, \bibinfo {author}
  {\bibfnamefont {W.}~\bibnamefont {Bao}}, \ and\ \bibinfo {author}
  {\bibfnamefont {D.}~\bibnamefont {Jaksch}},\ }\href {\doibase
  10.1209/0295-5075/98/26001} {\bibfield  {journal} {\bibinfo  {journal} {EPL}\
  }\textbf {\bibinfo {volume} {98}},\ \bibinfo {pages} {26001} (\bibinfo {year}
  {2012})}\BibitemShut {NoStop}%
\bibitem [{\citenamefont {Mistakidis}\ \emph {et~al.}(2019)\citenamefont
  {Mistakidis}, \citenamefont {Katsimiga}, \citenamefont {Koutentakis},
  \citenamefont {Busch},\ and\ \citenamefont {Schmelcher}}]{Mistakidis2019a}%
  \BibitemOpen
  \bibfield  {author} {\bibinfo {author} {\bibfnamefont {S.~I.}\ \bibnamefont
  {Mistakidis}}, \bibinfo {author} {\bibfnamefont {G.~C.}\ \bibnamefont
  {Katsimiga}}, \bibinfo {author} {\bibfnamefont {G.~M.}\ \bibnamefont
  {Koutentakis}}, \bibinfo {author} {\bibfnamefont {T.}~\bibnamefont {Busch}},
  \ and\ \bibinfo {author} {\bibfnamefont {P.}~\bibnamefont {Schmelcher}},\
  }\href {\doibase 10.1103/PhysRevLett.122.183001} {\bibfield  {journal}
  {\bibinfo  {journal} {Phys. Rev. Lett.}\ }\textbf {\bibinfo {volume} {122}},\
  \bibinfo {pages} {183001} (\bibinfo {year} {2019})}\BibitemShut {NoStop}%
\bibitem [{\citenamefont {Mistakidis}\ \emph {et~al.}(2021)\citenamefont
  {Mistakidis}, \citenamefont {Koutentakis}, \citenamefont {Grusdt},
  \citenamefont {Sadeghpour},\ and\ \citenamefont
  {Schmelcher}}]{Mistakidis2021a}%
  \BibitemOpen
  \bibfield  {author} {\bibinfo {author} {\bibfnamefont {S.~I.}\ \bibnamefont
  {Mistakidis}}, \bibinfo {author} {\bibfnamefont {G.~M.}\ \bibnamefont
  {Koutentakis}}, \bibinfo {author} {\bibfnamefont {F.}~\bibnamefont {Grusdt}},
  \bibinfo {author} {\bibfnamefont {H.~R.}\ \bibnamefont {Sadeghpour}}, \ and\
  \bibinfo {author} {\bibfnamefont {P.}~\bibnamefont {Schmelcher}},\ }\href
  {\doibase 10.1088/1367-2630/abe9d5} {\bibfield  {journal} {\bibinfo
  {journal} {New J. Phys.}\ }\textbf {\bibinfo {volume} {23}},\ \bibinfo
  {pages} {043051} (\bibinfo {year} {2021})}\BibitemShut {NoStop}%
\bibitem [{\citenamefont {Massignan}\ \emph {et~al.}(2014)\citenamefont
  {Massignan}, \citenamefont {Zaccanti},\ and\ \citenamefont
  {Bruun}}]{Massignan2014Polarons}%
  \BibitemOpen
  \bibfield  {author} {\bibinfo {author} {\bibfnamefont {P.}~\bibnamefont
  {Massignan}}, \bibinfo {author} {\bibfnamefont {M.}~\bibnamefont {Zaccanti}},
  \ and\ \bibinfo {author} {\bibfnamefont {G.~M.}\ \bibnamefont {Bruun}},\
  }\href {\doibase 10.1088/0034-4885/77/3/034401} {\bibfield  {journal}
  {\bibinfo  {journal} {Rep. Prog. Phys.}\ }\textbf {\bibinfo {volume} {77}},\
  \bibinfo {pages} {034401} (\bibinfo {year} {2014})}\BibitemShut {NoStop}%
\bibitem [{\citenamefont {Mertes}\ \emph {et~al.}(2007)\citenamefont {Mertes},
  \citenamefont {Merrill}, \citenamefont {{Carretero-Gonz{\'a}lez}},
  \citenamefont {Frantzeskakis}, \citenamefont {Kevrekidis},\ and\
  \citenamefont {Hall}}]{Mertes2007}%
  \BibitemOpen
  \bibfield  {author} {\bibinfo {author} {\bibfnamefont {K.~M.}\ \bibnamefont
  {Mertes}}, \bibinfo {author} {\bibfnamefont {J.~W.}\ \bibnamefont {Merrill}},
  \bibinfo {author} {\bibfnamefont {R.}~\bibnamefont
  {{Carretero-Gonz{\'a}lez}}}, \bibinfo {author} {\bibfnamefont {D.~J.}\
  \bibnamefont {Frantzeskakis}}, \bibinfo {author} {\bibfnamefont {P.~G.}\
  \bibnamefont {Kevrekidis}}, \ and\ \bibinfo {author} {\bibfnamefont {D.~S.}\
  \bibnamefont {Hall}},\ }\href {\doibase 10.1103/PhysRevLett.99.190402}
  {\bibfield  {journal} {\bibinfo  {journal} {Phys. Rev. Lett.}\ }\textbf
  {\bibinfo {volume} {99}},\ \bibinfo {pages} {190402} (\bibinfo {year}
  {2007})}\BibitemShut {NoStop}%
\bibitem [{\citenamefont {Pollack}\ \emph {et~al.}(2009)\citenamefont
  {Pollack}, \citenamefont {Dries}, \citenamefont {Junker}, \citenamefont
  {Chen}, \citenamefont {Corcovilos},\ and\ \citenamefont
  {Hulet}}]{Pollack2009}%
  \BibitemOpen
  \bibfield  {author} {\bibinfo {author} {\bibfnamefont {S.~E.}\ \bibnamefont
  {Pollack}}, \bibinfo {author} {\bibfnamefont {D.}~\bibnamefont {Dries}},
  \bibinfo {author} {\bibfnamefont {M.}~\bibnamefont {Junker}}, \bibinfo
  {author} {\bibfnamefont {Y.~P.}\ \bibnamefont {Chen}}, \bibinfo {author}
  {\bibfnamefont {T.~A.}\ \bibnamefont {Corcovilos}}, \ and\ \bibinfo {author}
  {\bibfnamefont {R.~G.}\ \bibnamefont {Hulet}},\ }\href {\doibase
  10.1103/PhysRevLett.102.090402} {\bibfield  {journal} {\bibinfo  {journal}
  {Phys. Rev. Lett.}\ }\textbf {\bibinfo {volume} {102}},\ \bibinfo {pages}
  {090402} (\bibinfo {year} {2009})}\BibitemShut {NoStop}%
\bibitem [{\citenamefont {Charalampidis}\ \emph
  {et~al.}(2018{\natexlab{b}})\citenamefont {Charalampidis}, \citenamefont
  {Lee}, \citenamefont {Kevrekidis},\ and\ \citenamefont
  {Chong}}]{Charalampidis2018a}%
  \BibitemOpen
  \bibfield  {author} {\bibinfo {author} {\bibfnamefont {E.~G.}\ \bibnamefont
  {Charalampidis}}, \bibinfo {author} {\bibfnamefont {J.}~\bibnamefont {Lee}},
  \bibinfo {author} {\bibfnamefont {P.~G.}\ \bibnamefont {Kevrekidis}}, \ and\
  \bibinfo {author} {\bibfnamefont {C.}~\bibnamefont {Chong}},\ }\href
  {\doibase 10.1103/PhysRevE.98.032903} {\bibfield  {journal} {\bibinfo
  {journal} {Phys. Rev. E}\ }\textbf {\bibinfo {volume} {98}},\ \bibinfo
  {pages} {032903} (\bibinfo {year} {2018}{\natexlab{b}})}\BibitemShut
  {NoStop}%
\bibitem [{\citenamefont {Miyazawa}\ \emph {et~al.}(2021)\citenamefont
  {Miyazawa}, \citenamefont {Chong}, \citenamefont {Kevrekidis},\ and\
  \citenamefont {Yang}}]{Miyazawa2021}%
  \BibitemOpen
  \bibfield  {author} {\bibinfo {author} {\bibfnamefont {Y.}~\bibnamefont
  {Miyazawa}}, \bibinfo {author} {\bibfnamefont {C.}~\bibnamefont {Chong}},
  \bibinfo {author} {\bibfnamefont {P.~G.}\ \bibnamefont {Kevrekidis}}, \ and\
  \bibinfo {author} {\bibfnamefont {J.}~\bibnamefont {Yang}},\ }\href@noop {}
  {\bibfield  {journal} {\bibinfo  {journal} {ArXiv210713169 Nlin}\ } (\bibinfo
  {year} {2021})},\ \Eprint {http://arxiv.org/abs/2107.13169} {arXiv:2107.13169
  [nlin]} \BibitemShut {NoStop}%
\bibitem [{\citenamefont {Biondini}\ and\ \citenamefont
  {Oregero}(2020)}]{Biondini2020Semiclassical}%
  \BibitemOpen
  \bibfield  {author} {\bibinfo {author} {\bibfnamefont {G.}~\bibnamefont
  {Biondini}}\ and\ \bibinfo {author} {\bibfnamefont {J.}~\bibnamefont
  {Oregero}},\ }\href {\doibase 10.1111/sapm.12321} {\bibfield  {journal}
  {\bibinfo  {journal} {Stud. Appl. Math.}\ }\textbf {\bibinfo {volume}
  {145}},\ \bibinfo {pages} {325} (\bibinfo {year} {2020})}\BibitemShut
  {NoStop}%
\bibitem [{Note2()}]{Note2}%
  \BibitemOpen
  \bibinfo {note} {For a transverse confinement with $\omega _{\perp }=2\pi
  \times 150\protect \tmspace +\thinmuskip {.1667em}$Hz these widths refer to a
  range from $45\protect \tmspace +\thinmuskip {.1667em}\mu $m to $135\protect
  \tmspace +\thinmuskip {.1667em}\mu $m.}\BibitemShut {Stop}%
\bibitem [{\citenamefont {Kevrekidis}\ and\ \citenamefont
  {Frantzeskakis}(2016)}]{Kevrekidis2016Solitons}%
  \BibitemOpen
  \bibfield  {author} {\bibinfo {author} {\bibfnamefont {P.~G.}\ \bibnamefont
  {Kevrekidis}}\ and\ \bibinfo {author} {\bibfnamefont {D.~J.}\ \bibnamefont
  {Frantzeskakis}},\ }\href {\doibase 10.1016/j.revip.2016.07.002} {\bibfield
  {journal} {\bibinfo  {journal} {Rev. Phys.}\ }\textbf {\bibinfo {volume}
  {1}},\ \bibinfo {pages} {140} (\bibinfo {year} {2016})}\BibitemShut {NoStop}%
\bibitem [{\citenamefont {Busch}\ and\ \citenamefont
  {Anglin}(2001)}]{Busch2001}%
  \BibitemOpen
  \bibfield  {author} {\bibinfo {author} {\bibfnamefont {T.}~\bibnamefont
  {Busch}}\ and\ \bibinfo {author} {\bibfnamefont {J.~R.}\ \bibnamefont
  {Anglin}},\ }\href {\doibase 10.1103/PhysRevLett.87.010401} {\bibfield
  {journal} {\bibinfo  {journal} {Phys. Rev. Lett.}\ }\textbf {\bibinfo
  {volume} {87}},\ \bibinfo {pages} {010401} (\bibinfo {year}
  {2001})}\BibitemShut {NoStop}%
\bibitem [{\citenamefont {Yan}\ \emph {et~al.}(2014)\citenamefont {Yan},
  \citenamefont {{Carretero-Gonz{\'a}lez}}, \citenamefont {Frantzeskakis},
  \citenamefont {Kevrekidis}, \citenamefont {Proukakis},\ and\ \citenamefont
  {Spirn}}]{Yan2014Exploring}%
  \BibitemOpen
  \bibfield  {author} {\bibinfo {author} {\bibfnamefont {D.}~\bibnamefont
  {Yan}}, \bibinfo {author} {\bibfnamefont {R.}~\bibnamefont
  {{Carretero-Gonz{\'a}lez}}}, \bibinfo {author} {\bibfnamefont {D.~J.}\
  \bibnamefont {Frantzeskakis}}, \bibinfo {author} {\bibfnamefont {P.~G.}\
  \bibnamefont {Kevrekidis}}, \bibinfo {author} {\bibfnamefont {N.~P.}\
  \bibnamefont {Proukakis}}, \ and\ \bibinfo {author} {\bibfnamefont
  {D.}~\bibnamefont {Spirn}},\ }\href {\doibase 10.1103/PhysRevA.89.043613}
  {\bibfield  {journal} {\bibinfo  {journal} {Phys. Rev. A}\ }\textbf {\bibinfo
  {volume} {89}},\ \bibinfo {pages} {043613} (\bibinfo {year}
  {2014})}\BibitemShut {NoStop}%
\bibitem [{\citenamefont {Cockburn}\ \emph {et~al.}(2010)\citenamefont
  {Cockburn}, \citenamefont {Nistazakis}, \citenamefont {Horikis},
  \citenamefont {Kevrekidis}, \citenamefont {Proukakis},\ and\ \citenamefont
  {Frantzeskakis}}]{Cockburn2010MatterWave}%
  \BibitemOpen
  \bibfield  {author} {\bibinfo {author} {\bibfnamefont {S.~P.}\ \bibnamefont
  {Cockburn}}, \bibinfo {author} {\bibfnamefont {H.~E.}\ \bibnamefont
  {Nistazakis}}, \bibinfo {author} {\bibfnamefont {T.~P.}\ \bibnamefont
  {Horikis}}, \bibinfo {author} {\bibfnamefont {P.~G.}\ \bibnamefont
  {Kevrekidis}}, \bibinfo {author} {\bibfnamefont {N.~P.}\ \bibnamefont
  {Proukakis}}, \ and\ \bibinfo {author} {\bibfnamefont {D.~J.}\ \bibnamefont
  {Frantzeskakis}},\ }\href {\doibase 10.1103/PhysRevLett.104.174101}
  {\bibfield  {journal} {\bibinfo  {journal} {Phys. Rev. Lett.}\ }\textbf
  {\bibinfo {volume} {104}},\ \bibinfo {pages} {174101} (\bibinfo {year}
  {2010})}\BibitemShut {NoStop}%
\bibitem [{\citenamefont {Cockburn}\ and\ \citenamefont
  {Proukakis}(2009)}]{Cockburn2009stochastic}%
  \BibitemOpen
  \bibfield  {author} {\bibinfo {author} {\bibfnamefont {S.~P.}\ \bibnamefont
  {Cockburn}}\ and\ \bibinfo {author} {\bibfnamefont {N.~P.}\ \bibnamefont
  {Proukakis}},\ }\href {\doibase 10.1134/S1054660X09040057} {\bibfield
  {journal} {\bibinfo  {journal} {Laser Phys.}\ }\textbf {\bibinfo {volume}
  {19}},\ \bibinfo {pages} {558} (\bibinfo {year} {2009})}\BibitemShut
  {NoStop}%
\bibitem [{\citenamefont {Bersano}\ \emph {et~al.}(2018)\citenamefont
  {Bersano}, \citenamefont {Gokhroo}, \citenamefont {Khamehchi}, \citenamefont
  {D'Ambroise}, \citenamefont {Frantzeskakis}, \citenamefont {Engels},\ and\
  \citenamefont {Kevrekidis}}]{Bersano2018ThreeComponent}%
  \BibitemOpen
  \bibfield  {author} {\bibinfo {author} {\bibfnamefont {T.~M.}\ \bibnamefont
  {Bersano}}, \bibinfo {author} {\bibfnamefont {V.}~\bibnamefont {Gokhroo}},
  \bibinfo {author} {\bibfnamefont {M.~A.}\ \bibnamefont {Khamehchi}}, \bibinfo
  {author} {\bibfnamefont {J.}~\bibnamefont {D'Ambroise}}, \bibinfo {author}
  {\bibfnamefont {D.~J.}\ \bibnamefont {Frantzeskakis}}, \bibinfo {author}
  {\bibfnamefont {P.}~\bibnamefont {Engels}}, \ and\ \bibinfo {author}
  {\bibfnamefont {P.~G.}\ \bibnamefont {Kevrekidis}},\ }\href {\doibase
  10.1103/PhysRevLett.120.063202} {\bibfield  {journal} {\bibinfo  {journal}
  {Phys. Rev. Lett.}\ }\textbf {\bibinfo {volume} {120}},\ \bibinfo {pages}
  {063202} (\bibinfo {year} {2018})}\BibitemShut {NoStop}%
\bibitem [{\citenamefont {Katsimiga}\ \emph {et~al.}(2020)\citenamefont
  {Katsimiga}, \citenamefont {Mistakidis}, \citenamefont {Bersano},
  \citenamefont {Ome}, \citenamefont {Mossman}, \citenamefont {Mukherjee},
  \citenamefont {Schmelcher}, \citenamefont {Engels},\ and\ \citenamefont
  {Kevrekidis}}]{Katsimiga2020}%
  \BibitemOpen
  \bibfield  {author} {\bibinfo {author} {\bibfnamefont {G.~C.}\ \bibnamefont
  {Katsimiga}}, \bibinfo {author} {\bibfnamefont {S.~I.}\ \bibnamefont
  {Mistakidis}}, \bibinfo {author} {\bibfnamefont {T.~M.}\ \bibnamefont
  {Bersano}}, \bibinfo {author} {\bibfnamefont {M.~K.~H.}\ \bibnamefont {Ome}},
  \bibinfo {author} {\bibfnamefont {S.~M.}\ \bibnamefont {Mossman}}, \bibinfo
  {author} {\bibfnamefont {K.}~\bibnamefont {Mukherjee}}, \bibinfo {author}
  {\bibfnamefont {P.}~\bibnamefont {Schmelcher}}, \bibinfo {author}
  {\bibfnamefont {P.}~\bibnamefont {Engels}}, \ and\ \bibinfo {author}
  {\bibfnamefont {P.~G.}\ \bibnamefont {Kevrekidis}},\ }\href {\doibase
  10.1103/PhysRevA.102.023301} {\bibfield  {journal} {\bibinfo  {journal}
  {Phys. Rev. A}\ }\textbf {\bibinfo {volume} {102}},\ \bibinfo {pages}
  {023301} (\bibinfo {year} {2020})}\BibitemShut {NoStop}%
\bibitem [{\citenamefont {Katsimiga}\ \emph {et~al.}(2017)\citenamefont
  {Katsimiga}, \citenamefont {Koutentakis}, \citenamefont {Mistakidis},
  \citenamefont {Kevrekidis},\ and\ \citenamefont
  {Schmelcher}}]{Katsimiga2017}%
  \BibitemOpen
  \bibfield  {author} {\bibinfo {author} {\bibfnamefont {G.~C.}\ \bibnamefont
  {Katsimiga}}, \bibinfo {author} {\bibfnamefont {G.~M.}\ \bibnamefont
  {Koutentakis}}, \bibinfo {author} {\bibfnamefont {S.~I.}\ \bibnamefont
  {Mistakidis}}, \bibinfo {author} {\bibfnamefont {P.~G.}\ \bibnamefont
  {Kevrekidis}}, \ and\ \bibinfo {author} {\bibfnamefont {P.}~\bibnamefont
  {Schmelcher}},\ }\href {\doibase 10.1088/1367-2630/aa766b} {\bibfield
  {journal} {\bibinfo  {journal} {New J. Phys.}\ }\textbf {\bibinfo {volume}
  {19}},\ \bibinfo {pages} {073004} (\bibinfo {year} {2017})}\BibitemShut
  {NoStop}%
\bibitem [{\citenamefont {Katsimiga}\ \emph {et~al.}(2018)\citenamefont
  {Katsimiga}, \citenamefont {Mistakidis}, \citenamefont {Koutentakis},
  \citenamefont {Kevrekidis},\ and\ \citenamefont
  {Schmelcher}}]{Katsimiga2018}%
  \BibitemOpen
  \bibfield  {author} {\bibinfo {author} {\bibfnamefont {G.~C.}\ \bibnamefont
  {Katsimiga}}, \bibinfo {author} {\bibfnamefont {S.~I.}\ \bibnamefont
  {Mistakidis}}, \bibinfo {author} {\bibfnamefont {G.~M.}\ \bibnamefont
  {Koutentakis}}, \bibinfo {author} {\bibfnamefont {P.~G.}\ \bibnamefont
  {Kevrekidis}}, \ and\ \bibinfo {author} {\bibfnamefont {P.}~\bibnamefont
  {Schmelcher}},\ }\href {\doibase 10.1103/PhysRevA.98.013632} {\bibfield
  {journal} {\bibinfo  {journal} {Phys. Rev. A}\ }\textbf {\bibinfo {volume}
  {98}},\ \bibinfo {pages} {013632} (\bibinfo {year} {2018})}\BibitemShut
  {NoStop}%
\bibitem [{\citenamefont {Mistakidis}\ \emph {et~al.}(2018)\citenamefont
  {Mistakidis}, \citenamefont {Katsimiga}, \citenamefont {Kevrekidis},\ and\
  \citenamefont {Schmelcher}}]{Mistakidis2018}%
  \BibitemOpen
  \bibfield  {author} {\bibinfo {author} {\bibfnamefont {S.~I.}\ \bibnamefont
  {Mistakidis}}, \bibinfo {author} {\bibfnamefont {G.~C.}\ \bibnamefont
  {Katsimiga}}, \bibinfo {author} {\bibfnamefont {P.~G.}\ \bibnamefont
  {Kevrekidis}}, \ and\ \bibinfo {author} {\bibfnamefont {P.}~\bibnamefont
  {Schmelcher}},\ }\href {\doibase 10.1088/1367-2630/aabc6a} {\bibfield
  {journal} {\bibinfo  {journal} {New J. Phys.}\ }\textbf {\bibinfo {volume}
  {20}},\ \bibinfo {pages} {043052} (\bibinfo {year} {2018})}\BibitemShut
  {NoStop}%
\bibitem [{\citenamefont {El'}\ \emph {et~al.}(1993)\citenamefont {El'},
  \citenamefont {Gurevich}, \citenamefont {Khodorovski{\v i}},\ and\
  \citenamefont {Krylov}}]{El1993}%
  \BibitemOpen
  \bibfield  {author} {\bibinfo {author} {\bibfnamefont {G.~A.}\ \bibnamefont
  {El'}}, \bibinfo {author} {\bibfnamefont {A.~V.}\ \bibnamefont {Gurevich}},
  \bibinfo {author} {\bibfnamefont {V.~V.}\ \bibnamefont {Khodorovski{\v i}}},
  \ and\ \bibinfo {author} {\bibfnamefont {A.~L.}\ \bibnamefont {Krylov}},\
  }\href {\doibase 10.1016/0375-9601(93)90015-R} {\bibfield  {journal}
  {\bibinfo  {journal} {Physics Letters A}\ }\textbf {\bibinfo {volume}
  {177}},\ \bibinfo {pages} {357} (\bibinfo {year} {1993})}\BibitemShut
  {NoStop}%
\bibitem [{\citenamefont {Zakharov}\ and\ \citenamefont
  {Gelash}(2013)}]{Zakharov2013}%
  \BibitemOpen
  \bibfield  {author} {\bibinfo {author} {\bibfnamefont {V.~E.}\ \bibnamefont
  {Zakharov}}\ and\ \bibinfo {author} {\bibfnamefont {A.~A.}\ \bibnamefont
  {Gelash}},\ }\href {\doibase 10.1103/PhysRevLett.111.054101} {\bibfield
  {journal} {\bibinfo  {journal} {Phys. Rev. Lett.}\ }\textbf {\bibinfo
  {volume} {111}},\ \bibinfo {pages} {054101} (\bibinfo {year}
  {2013})}\BibitemShut {NoStop}%
\bibitem [{\citenamefont {Biondini}(2018)}]{Biondini2018a}%
  \BibitemOpen
  \bibfield  {author} {\bibinfo {author} {\bibfnamefont {G.}~\bibnamefont
  {Biondini}},\ }\href {\doibase 10.1103/PhysRevE.98.052220} {\bibfield
  {journal} {\bibinfo  {journal} {Phys. Rev. E}\ }\textbf {\bibinfo {volume}
  {98}},\ \bibinfo {pages} {052220} (\bibinfo {year} {2018})}\BibitemShut
  {NoStop}%
\bibitem [{\citenamefont {Akhmediev}\ \emph {et~al.}(1987)\citenamefont
  {Akhmediev}, \citenamefont {Eleonskii},\ and\ \citenamefont
  {Kulagin}}]{Akhmediev1987}%
  \BibitemOpen
  \bibfield  {author} {\bibinfo {author} {\bibfnamefont {N.~N.}\ \bibnamefont
  {Akhmediev}}, \bibinfo {author} {\bibfnamefont {V.~M.}\ \bibnamefont
  {Eleonskii}}, \ and\ \bibinfo {author} {\bibfnamefont {N.~E.}\ \bibnamefont
  {Kulagin}},\ }\href {\doibase 10.1007/BF01017105} {\bibfield  {journal}
  {\bibinfo  {journal} {Theor Math Phys}\ }\textbf {\bibinfo {volume} {72}},\
  \bibinfo {pages} {809} (\bibinfo {year} {1987})}\BibitemShut {NoStop}%
\bibitem [{\citenamefont {Chen}\ and\ \citenamefont {Hung}(2021)}]{Chen2021}%
  \BibitemOpen
  \bibfield  {author} {\bibinfo {author} {\bibfnamefont {C.-A.}\ \bibnamefont
  {Chen}}\ and\ \bibinfo {author} {\bibfnamefont {C.-L.}\ \bibnamefont
  {Hung}},\ }\href {\doibase 10.1103/PhysRevLett.127.023604} {\bibfield
  {journal} {\bibinfo  {journal} {Phys. Rev. Lett.}\ }\textbf {\bibinfo
  {volume} {127}},\ \bibinfo {pages} {023604} (\bibinfo {year}
  {2021})}\BibitemShut {NoStop}%
\end{thebibliography}%

\end{document}